\documentclass[pre, reprint, showpacs, superscriptaddress]{revtex4-1}
\usepackage{amsmath, graphicx,amssymb,color, bm}
\newcommand{\colorred }{\color{black}}

\begin{document}
\title{Nucleation of symmetric domains in the coupled leaflets of a bilayer}

\date{\today}

\author{J.~J.~Williamson}
\email{johnjosephwilliamson@gmail.com}
\author{P.~D.~Olmsted}
\email{pdo7@georgetown.edu}
\affiliation{Department of Physics, Institute for Soft Matter Synthesis and Metrology, Georgetown University, 37th and O Streets, N.W., Washington, D.C. 20057, USA}

\begin{abstract} 
We study the kinetics governing the attainment of inter-leaflet domain symmetry in a phase-separating amphiphilic bilayer. ``Indirect'' inter-leaflet coupling via hydrophobic mismatch can induce an instability towards a metastable pattern of locally asymmetric domains upon quenching from high temperature. This necessitates a nucleation step to form the conventional symmetric pattern of domains, which are favoured by a ``direct'' inter-leaflet coupling. We model the energetics for a symmetric domain to nucleate from the metastable state, and find that an interplay between hydrophobic mismatch and thickness stretching/compression causes the effective hydrophobic mismatch, and thus line tension, to depend on domain size. This leads to strong departure from classical nucleation theory. We speculate on implications for cell membrane rafts or clusters, whose size may be of similar magnitude to estimated critical radii for domain symmetry. 
\end{abstract}

\maketitle

\section{\label{sec:intro}Introduction}

\begin{figure}[floatfix]
\includegraphics[width=8cm]{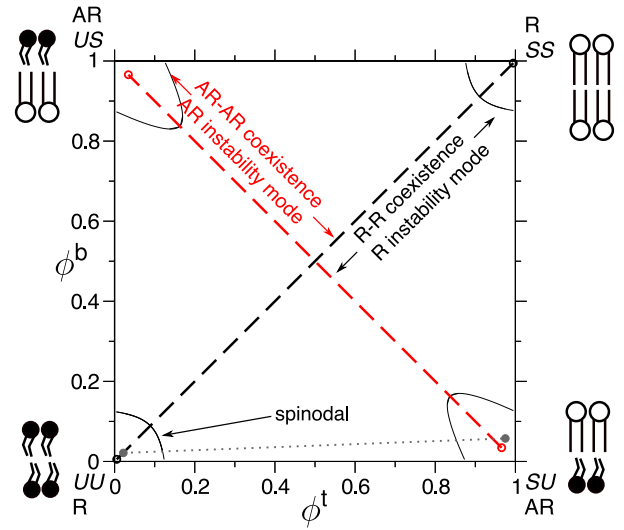}
\caption{\colorred \label{schematic}Partial phase diagram showing competing equilibrium R-R (black) and metastable AR-AR (red) phase coexistences \cite{Williamson2014}. Spinodals enclose the regions of local stability. Cartoons of the dominant inter-leaflet arrangement in each bilayer phase are shown. The grey dotted line illustrates R-AR coexistence, which is briefly discussed in Section~\ref{sec:def}. Other phase coexistences not considered here are omitted \cite{Williamson2014}. Parameters:\ $\Delta_0 = 2\,a$, $\kappa = 3\,a^{-2}k_\textrm{B}T$, $V=0.6\,k_\textrm{B}T$, $J = 4\,a^{-2}k_\textrm{B}T$, $B=0.48\,a^{-2}k_\textrm{B}T$.}
\end{figure}

A phase-separating lipid (or other amphiphilic) bilayer may access competing equilibrium and metastable phase coexistences, due to the presence of two leaflets subject to competing inter-leaflet couplings (Figs.~\ref{schematic}, \ref{lattice}a) \cite{Williamson2014, Williamson2015}. A ``direct" inter-leaflet coupling \cite{Putzel2008, Putzel2011, May2009, Collins2008, Risselada2008} promotes registered (R) bilayer phases, in which both leaflets are locally dominated by the same species.
An ``indirect'' coupling from hydrophobic tail length mismatch favours uniform bilayer thickness \cite{Stevens2005, Perlmutter2011, Reigada2015, Zhang2004, Reigada2015} and thus promotes antiregistered (AR) phases, in which the leaflets are dominated by different species so that the bilayer is locally compositionally asymmetric.
Depending on the choice of overall leaflet compositions, two (e.g., R-R or AR-AR) or three bilayer phases may coexist. \colorred 
For example, asymmetric overall leaflet compositions can lead to two approximately symmetric bilayer phases and a highly asymmetric one (R-R-AR), first observed and explained in \cite{Collins2008}. Phase equilibria of coupled bilayer leaflets were subsequently studied using phenomenological free energies \cite{Wagner2007, Putzel2008, May2009}. However, a full description of the indirect coupling described above requires a model that microscopically incorporates hydrophobic mismatch \cite{Williamson2014}. \color{black}

Using such a model, we have shown how coexistence of antiregistered phases can be kinetically preferred due to the effect of hydrophobic mismatch \cite{Williamson2014} so that, before equilibrating, a quenched bilayer must escape a (typically metastable) locally asymmetric state. The understanding of this novel statistical thermodynamics will allow greater control over artificial membranes, and the underlying interactions are expected to play a role in transmembrane organisation of rafts or clusters \textit{in vivo}, with possible relevance to signalling \cite{Kusumi2004} and anaesthetic action \cite{Reigada2015}.
\colorred The predicted behaviour constitutes an example of Ostwald's ``rule of stages'' \cite{Ostwald}, by which a system will pass through available metastable states on its way to equilibrium. Ostwald's rule is familiar (via different origins) in colloids, metallurgy and drug design.
\color{black}

Immediately after quenching a bilayer to a phase-separating region of parameter space, any spinodal instabilities to which the uniform state is subject compete to determine the dominant initial demixing mode. A bilayer with roughly equimolar composition in each leaflet \footnote{In a real system in which lipid species' molecular areas may differ, $\Phi^\textrm{t}\!=\!\Phi^\textrm{b}\!=\!0.5$ refers instead to an equal area fractions mixture in each leaflet.} can be subject both to an ``R mode'' with composition perturbations locally symmetric between leaflets, and a perpendicular AR mode with asymmetric perturbations (Fig.~\ref{schematic}). If the AR mode is fastest-growing, spinodal decomposition to AR-AR coexistence occurs first, leading to local asymmetry throughout the bilayer.
To equilibrate to R-R from a metastable AR-AR state, the bilayer must undergo nucleation of registered bilayer phases.
Hence, three classes of kinetics arise:\ direct separation into equilibrium phases, equilibration via nucleation out of a metastable state, or trapping in a metastable state. For other overall compositions subject to competing instabilities, the competition of symmetric and asymmetric phases is qualitatively similar \cite{Williamson2015}, though more complex. 

In this paper, we focus on the nucleation energetics that determine whether equilibrium domain symmetry is reached from a metastable state. \colorred First, we introduce the model and discuss the interpretation of bilayer domain symmetry and asymmetry via phase diagrams with a composition axis for each leaflet, with reference to existing experiment and theory. \color{black}We then identify the three classes of kinetics in simulation, guided by a linear instability analysis, and develop a theory for the nucleation of registered domains, which captures the interplay of bulk free energy with thickness mismatch occurring at the perimeter of a registered domain. Together with the linear stability analysis, the calculated nucleation energetics are consistent with the simulation results. We find that the effective hydrophobic mismatch between a domain and its surroundings is domain size-dependent, which causes strong departure from classical nucleation theory. 

\section{Model}

\colorred A detailed description of the model appears in \cite{Williamson2014, Williamson2015}. We briefly recapitulate the model and its analysis in terms of phase diagrams and kinetics.
A schematic of the lattice model and how it is coarse-grained is shown in Fig.~\ref{lattice}.
\color{black}

\subsection{Lattice model}\label{sec:lattice}

\begin{figure}[floatfix]
\includegraphics[width=8cm]{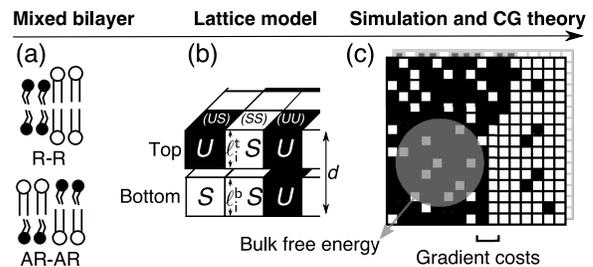}
\caption{\colorred \label{lattice}(a)~Mixed bilayer containing $S$ and $U$  ($L_o$ or gel and $L_d$-like) model species, illustrating the locally symmetric (R-R) and locally asymmetric (AR-AR) phase coexistences considered here. 
(b)~Microscopic lattice model for coupled leaflets, which can be coarse-grained (CG) (c) to give the mean-field free-energy density $f(\phi^\textrm{t},\phi^\textrm{b})$ as a function of locally-averaged leaflet compositions, and analysed for kinetics of domain formation with the inclusion of gradient costs for domain boundaries \cite{Williamson2014}. The lattice model can also be directly simulated.}
\end{figure}

A local bilayer patch is described as $N$ lattice sites with top (t) and bottom (b)-leaflet lipids (Fig.~\ref{lattice}b).
The lipids' hydrophobic lengths $\ell^\textrm{t(b)}_i$ lead to the total bilayer thickness 
$d_i \equiv \ell^\textrm{t}_i + \ell^\textrm{b}_i$
and inter-leaflet difference 
$
\Delta_i \equiv \ell^\textrm{t}_i - \ell^\textrm{b}_i
$.
Model species $S$ and $U$ represent saturated and unsaturated lipids or, e.g., the liquid-ordered ($L_{o}$) and liquid-disordered ($L_{d}$) states in a ternary mixture \cite{Williamson2015}.

The Hamiltonian is 
\begin{align}\label{eqn:fun4}
H = &\sum_{<i,j>} ( V_{\hat{\phi}_i^\textrm{t} \hat{\phi}_j^\textrm{t}}  +  V_{\hat{\phi}_i^\textrm{b} \hat{\phi}_j^\textrm{b}}) 
+ \sum_{<i,j>} \tfrac{1}{2}\tilde{J} (d_i - d_j)^2 \notag \\
&+  \sum_{i} \tfrac{1}{2}B (\Delta_i)^2 
+  \sum_{i} \tfrac{1}{2}\kappa \left(( \ell^\textrm{t}_i - \ell_0^{\textrm{t}i})^2 + ( \ell^\textrm{b}_i - 
\ell_0^{\textrm{b}i})^2                      \right)~,
\end{align}
\noindent where $\hat{\phi}^\textrm{t(b)}_i = 1$ if the top (bottom) of lattice site $i$ contains an $S$ lipid, $\hat{\phi}^\textrm{t(b)}_i = 0$ if $U$. The species-dependent ideal (i.e., preferred) hydrophobic tail lengths are $ \ell_0^{\textrm{t(b)}i} = \ell_{S0}$ for an $S$ lipid at the top (bottom) of site $i$, or $\ell_{U0}$ for $U$, and each site is pairwise registered (R, $SS$ or $UU$) or antiregistered (AR, $SU$ or $US$). 

\color{black}
$V \equiv V_{10} - \tfrac{1}{2}(V_{00} + V_{11})$ quantifies purely intra-leaflet interactions, such as those between headgroups. 
The ``direct'' coupling $B$ promotes pairwise R between lipids, nominally by penalising tail structure mismatch (which we treat as implicit in tail length mismatch \cite{Komura2004}) across the midplane. The particular mechanisms responsible for the direct coupling are not crucial to our model, however -- for comparison with the literature we can simply estimate an effective strength of the conventional inter-leaflet mismatch energy $\gamma$ \cite{Williamson2014}, which is shown on Fig.~\ref{kinetic} as well \cite{Pantano2011, Risselada2008, Polley2013, Putzel2011, May2009, Blosser2015}.
The hydrophobic ``indirect'' coupling $\tilde{J}$ promotes pairwise AR, by penalising mismatch in the bilayer thickness profile. We also define $J \equiv 4\tilde{J}$, which appears in the mean-field approximation of Eq.~\ref{eqn:fun4} used to derive the coarse-grained free energy.  
$\kappa$ can be related to the area compression modulus $\kappa_A$ \cite{Williamson2014}, and penalises variation from species-dependent ideal length. Weaker $\kappa$ means the species can more easily adapt their tail length and structure to one another's presence. The mismatch parameter $\Delta_0 \equiv \ell_{S0} - \ell_{U0}$ is cast as a length, but represents \textit{both} tail length and structure mismatch;\ it couples to both the indirect and direct inter-leaflet couplings, $J$ and $B$. Once fiducial values of the parameters are set, varying $J$ alone approximates changing the mismatch in tail length but not structure (e.g., adding carbons to one species \cite{Perlmutter2011}), while varying $B$ alone approximates varying the mismatch in tail structure, e.g., unsaturation.
We arbitrarily choose $\ell_{S0} > \ell_{U0}$. 
The reference total thickness $d_0 \equiv \ell_{S0} + \ell_{U0}$ is irrelevant in the absence of an external field acting on bilayer thickness. 

\color{black}

\subsection{Simulation protocol}

\colorred We simulate a $\mathcal{L}^2 = \mathcal{N}$ bilayer where $\mathcal{L} = 100$ (script letters refer to the entire simulated bilayer, as opposed to a local bilayer patch). The Kinetic Monte Carlo scheme \cite{Williamson2015} resembles Kawasaki (spin-exchange) dynamics, governed by the Hamiltonian (Eq.~\ref{eqn:fun4}):\ lipids exchange between neighbouring lattice sites within their leaflet, thus mimicking diffusive evolution after a quench from a high-temperature, randomised initial configuration. Hydrodynamics are not included, but we do not expect this to significantly alter the conclusions;\ we address this further in the Discussion.

\color{black}

The overall leaflet compositions are conserved, since we do not consider flip-flop or exchange with the solvent, and are given by
\begin{equation}\label{eqn:overallphi}
\Phi^\textrm{t(b)} \equiv \frac{\mathcal{N}_S^\textrm{t(b)} }{\mathcal{N}}~{\color{red},}
\end{equation}
\noindent 
\color{black}where $\mathcal{N}^\textrm{t(b)}_S$ is the total number of $S$ lipids in the top (bottom) leaflet. 
\color{black}R and AR lattice sites can interconvert ($SU + US \rightleftarrows SS + UU$) to alter the ``degree of registration'' (microscopic transbilayer symmetry) 
\begin{equation}
\lambda \equiv \frac{\mathcal{N}_{SS} + \mathcal{N}_{UU}}{\mathcal{N}}~,
\end{equation}
\noindent 
{\color{black}where $\mathcal{N}_{SS}$ is the total number of $SS$ lattice sites, etc.}

$\lambda$ can vary in the range
\begin{equation} \label{eqn:range}
\lvert \Phi^\textrm{t} +\Phi^\textrm{b} - 1\rvert \leq \lambda \leq 1 - \lvert \Phi^\textrm{t} - \Phi^\textrm{b} \rvert ~.
\end{equation}
Although the overall leaflet compositions are conserved, $\lambda$ is not \cite{Williamson2015}. Hence, the bilayer can phase-separate into either locally symmetric or asymmetric modes while the conserved \textit{overall} leaflet compositions may, as here, be fully symmetric between leaflets. 

Here we focus on symmetric, equimolar overall leaflet compositions $\Phi^\textrm{t}\!=\!\Phi^\textrm{b}\!=\!0.5$. By Eq.~\ref{eqn:range} the degree of registration $\lambda$ can vary from 0 (full pairwise antiregistration) to 1 (full pairwise registration).  

\subsection{Parameters}

The lattice spacing is $a \sim 0.8\, \textrm{nm}$. We use $\kappa = 3\,a^{-2}k_\textrm{B}T$, corresponding to $\kappa_{A} \approx 60\,k_\textrm{B}T\textrm{nm}^{-2}$, in the range for lipid bilayers at $300\,\textrm{K}$ \cite{Wallace2005, Needham1990, Rawicz2000}. A fiducial value of the indirect coupling parameter is $J \sim 2\,a^{-2}k_\textrm{B}T$ \cite{Williamson2014}, while $B$ leads to an effective value of $\gamma$ (Fig.~\ref{kinetic} secondary axis), for which existing estimates vary widely ($\gamma \sim 0.01 - 1\, k_\textrm{B}T\textrm{nm}^{-2}$ \cite{Pantano2011, Risselada2008, Polley2013, Putzel2011, May2009, Blosser2015}). 

\colorred We choose a mismatch parameter $\Delta_0 = 2\,a\!\sim\!1.6\,\textrm{nm}$, somewhat larger than the typical length mismatch in phospholipid mixtures, for which a (registered) phase thickness mismatch $\lesssim 2\,\textrm{nm}$ \cite{Garcia2007, Lin2006} would imply $\Delta_0 \lesssim 1\,\textrm{nm}$. Large $\Delta_0$ couples to both the indirect ($J$) and direct ($B$) couplings, increasing the energetic driving forces for both antiregistration and registration and making the competing phases clearer to interpret in simulation. The phenomenology is qualitatively similar upon reducing $\Delta_0$ \cite{Williamson2015}, with the caveat that $\Delta_0$ affects the effective value of $\gamma$ arising from a given $B$ (Eq.~\ref{eqn:mismatchenergy}). In Section~\ref{sec:nuc} we mention the dependence of the nucleation theory on $\Delta_0$. 
\color{black}

We use $V=0.6\,k_\textrm{B}T $ in the mean-field theory, above the mean-field threshold $V_0 \equiv 0.5\,k_\textrm{B}T$ for phase separation in the absence of other couplings, and use $V=0.9\, k_\textrm{B}T$ in simulation, where the corresponding threshold is $V_0^\textrm{sim.} = 0.88\,k_\textrm{B}T$ due to fluctuations \cite{Huang1987}. Although we cannot expect precise quantitative agreement between the mean-field theory and simulation, in this regime it appears the exact value of $V$ is not crucial to the kinetics;\ for example, a value $V = 0.9\,k_\textrm{B}T$ in the mean-field theory would yield a similar predicted landscape of relative R/AR growth rate $\Delta \omega$ to that calculated in Fig.~\ref{kinetic} \cite{Williamson2015}.

\section{Registered and antiregistered phases}\label{sec:comp}

\colorred 

We now discuss bilayer domain symmetry and asymmetry in terms of competing bilayer phase coexistences, focusing on those relevant to the present work and briefly considering those relevant to bilayers of asymmetric overall leaflet compositions ($\Phi^\textrm{b} \neq \Phi^\textrm{t}$).
Upon coarse-graining the microscopic model Eq.~\ref{eqn:fun4} (Fig.~\ref{lattice}c), a local bilayer patch is characterised by locally-averaged top and bottom leaflet compositions $\phi^\textrm{t(b)} \equiv \sum_i \hat{\phi}^\textrm{t(b)}_i/ N$. We derive a mean-field free-energy density $f(\phi^\textrm{t},\phi^\textrm{b})$ \cite{Williamson2014}, which yields a phase diagram in $(\phi^\textrm{t},\phi^\textrm{b})$ space (Fig.~\ref{schematic}) \cite{Williamson2014}. The coexisting bilayer phases determine the local order parameter in each leaflet, given by the projection of a given tie-line endpoint onto the $\phi^\textrm{t}$ or $\phi^\textrm{b}$ axis.

\subsection{Definition of registration and antiregistration} \label{sec:def}

Some experiments, particularly with asymmetric overall leaflet compositions, may use separate fluorophores to image domain morphology in each leaflet \cite{Garg2007, Visco2014, Lin2015}. Interpreting these in $(\phi^\textrm{t},\phi^\textrm{b})$ space, and relating them to experiments that do not image the separate leaflets \cite{Lin2006}, requires care. We define a registered (R) bilayer phase as one in which both leaflets are dominated by the same species. Hence, in our model, an R phase is dominated by $SS$ or $UU$ lattice sites (Fig.~\ref{schematic}) such that most lipids face one of their own species in the apposing leaflet. An antiregistered (AR) bilayer phase is one where the leaflets are dominated by opposite species. Hence, under our definition, ``registration'' is a property of a given homogeneous patch of the bilayer, describing approximate local compositional symmetry between the leaflets.

An alternative definition of registration is sometimes used, which we call ``colocalised enrichment'' \cite{Garg2007, Visco2014}. This describes a bilayer in which the regions of largest top-leaflet composition $\phi^\textrm{t}$ (relative to the average \textit{in that leaflet}) spatially superimpose on the regions of largest bottom-leaflet composition $\phi^\textrm{b}$. Colocalised enrichment is therefore a property of domain morphology over the entire bilayer, not of an individual bilayer phase. It requires a tie-line of \textit{finite positive slope} in $(\phi^\textrm{t},\phi^\textrm{b})$ space, so that:\ i)~both leaflets contain domains of larger and smaller than average $\phi^\textrm{t(b)}$;\ and ii)~the domains of large $\phi^\textrm{t}$ belong to the same bilayer phase as the domains of large $\phi^\textrm{b}$ and are thus spatially colocalised with them. 

R-AR coexistence (Fig.~\ref{schematic} dotted line) can be accessed by a bilayer of asymmetric overall leaflet compositions. R-AR tie-lines have positive slope, and thus exhibit colocalised enrichment, although the AR phase is highly asymmetric in composition. Hence, the colocalised enriched domains in each leaflet reported in \cite{Visco2014, Lin2015} (where the leaflets were separately imaged) are consistent with either an R-R or R-AR tie-line. Quantitative composition information would be required to unambiguously determine which was observed.

In the present work, we consider a bilayer of symmetric overall leaflet compositions for which R-R and AR-AR tie-lines compete. In this case, ``registration'' versus ``colocalised enrichment'' are practically equivalent. R-R has positive tie-line slope (colocalised enrichment), and both bilayer phases are compositionally symmetric (registered phases, under our definition). Conversely, AR-AR tie-lines are negatively-sloped so that enrichment in one leaflet colocalises with \textit{depletion} in the other, and both bilayer phases are compositionally asymmetric (antiregistered phases, under our definition).

The literature is ambiguous. For example, full colocalised enrichment (which can reflect either an R-R or R-AR tie-line) is described in \cite{Visco2014} as ``registration''. In contrast, in \cite{Reigada2015}, ``registration'' is also defined as the presence of $L_o$ (or $L_d$) on both sides of the bilayer, which is more like our definition of R phases versus AR phases. This ambiguity is not surprising because often, as here, the two definitions outlined above are similar. The two concepts can clash for asymmetric overall leaflet compositions \cite{Lin2015} where R-AR tie-lines can play a role.

In asymmetric supported bilayers in \cite{Visco2014}, the leaflets were imaged with separate fluorophores, revealing colocalised domains in both leaflets. The same was found in highly asymmetric vesicles in \cite{Lin2015}. On the face of it, these findings differ from some supported bilayers in \cite{Lin2006} which exhibited compositionally asymmetric regions interpreted as domains in \textit{only one leaflet}. These were inferred from variations in total bilayer thickness without imaging the separate leaflets. Height mismatch smaller than a known value for R-R was measured, indicating either R-AR or AR-AR-R coexistence. R-AR was then inferred by detecting that saturated lipids were predominantly in the top leaflet. The system was interpreted as having domains in only the top leaflet \cite{Lin2006}.

However, it is probable that the asymmetric bilayers in both \cite{Visco2014, Lin2015} and \cite{Lin2006} represent an R-AR tie-line. Thus, imaging the leaflets in \cite{Lin2006} separately could have shown colocalised enrichment just as in \cite{Visco2014, Lin2015}. The top-leaflet gel domains in \cite{Lin2006} may have apposed regions in the bottom leaflet weakly more ``gel-like'' than the average in the bottom leaflet. Similarly, the colocalised enrichment in \cite{Visco2014, Lin2015} does not require that bottom \cite{Visco2014} (or inner \cite{Lin2015}) leaflet domains are truly $L_o$ -- for R-AR they will, like the rest of the bottom/inner leaflet, be dominated by unsaturated lipids. It only requires that they are weakly more ``$L_o$-like'' than the average in their leaflet, i.e., that the R-AR tie-line is tilted. Hence, the question of whether domains in one leaflet ``induce domains in the other'' becomes the question of whether the given tie-line is tilted \textit{enough} for both leaflets to exhibit \textit{detectable} domain formation when a different fluorophore is used in each. The degree of R-AR tie-line tilt will depend on the direct inter-leaflet coupling $B$ and hence on molecular features.

\color{black}

\begin{figure*}[floatfix]
\includegraphics[width=16.0cm]{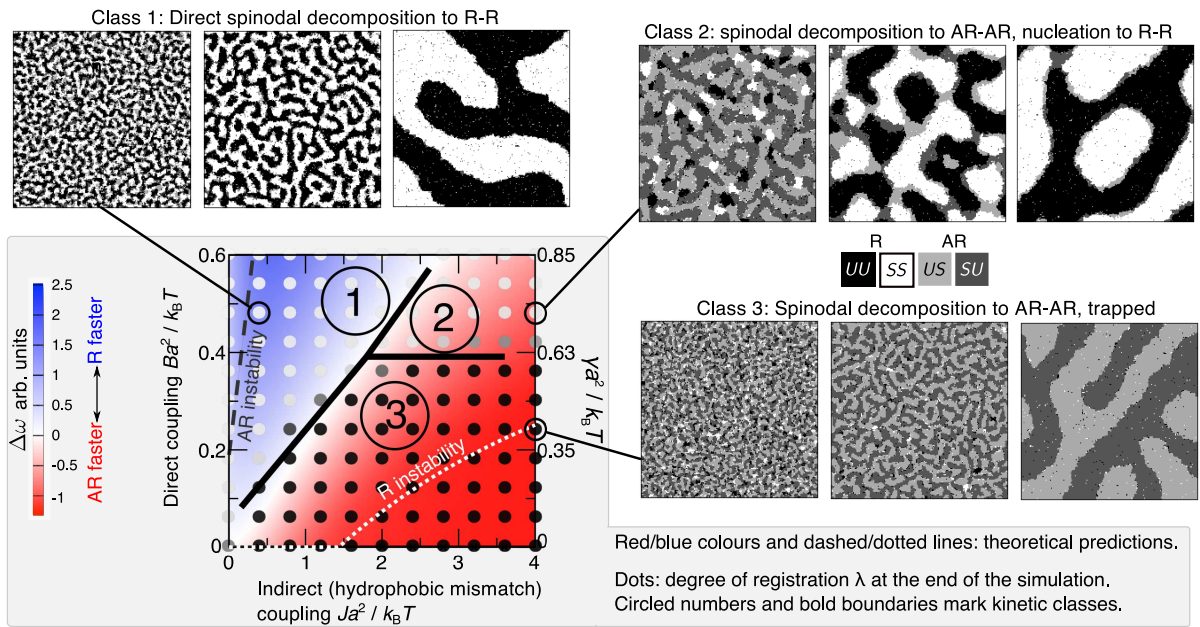}
\caption{\colorred \label{kinetic}Red/blue colours and dashed/dotted lines:\ mean-field theory parameter map showing relative growth rates of AR versus R instability modes for a bilayer comprising equimolar mixed leaflets, from linear stability analysis of initial demixing \cite{Williamson2014}, with $\Delta_0 = 2\,a$, $\kappa = 3\,a^{-2}k_\textrm{B}T$, $V=0.6\,k_\textrm{B}T$. The equilibrium state is R-R in all cases. Below the ``AR instability'' line AR-AR coexistence is possible but is metastable. Below the ``R instability'' line the homogeneous state is not unstable to the R mode, although R-R separation is still the equilibrium state.
Overlaid dots from simulation (with $V=0.9\,k_\textrm{B}T$) show the average degree of registration $\lambda$ at the end of the simulation time, $\lambda = 0$ (black, AR-AR) to $\lambda = 1$ (white, R-R). We identify three kinetic classes discussed in the text (circled 1, 2, 3, bold line marks approximate boundaries). Illustrative simulation snapshot sequences for each class are shown ($\mathcal{L}\!=\!200$ in the snapshots). Simulations are visualised with OVITO \cite{OVITO}.}
\end{figure*}

\subsection{Kinetics and competing phase coexistences}

For most typical parameter choices, the R phases are lower in $f(\phi^\textrm{t},\phi^\textrm{b})$ than are the AR, hence R-R tie-lines are equilibrium and AR-AR metastable \cite{Williamson2014}.
As well as the coarse-grained \textit{bulk} free energy $f(\phi^\textrm{t},\phi^\textrm{b})$, the underlying microscopic model yields free-energy costs for composition and thickness \textit{gradients}, which describe the role of spatial structure and domain formation in the kinetic competition of metastable and equilibrium states. Linear stability analysis of the initial homogeneous state (Fig.~\ref{kinetic}) predicts growth rates of competing R versus AR instability modes and thus whether symmetric or asymmetric domains form first \cite{Williamson2014}. The interplay between gradient and bulk free energies then governs the nucleation of symmetric domains (introduced in Section~\ref{sec:nuc}), which determines the eventual fate of a bilayer that has initially become metastably AR-AR. The ability to derive both gradient and bulk free energies from defined microscopic interactions is a key advantage of the present model over a purely phenomenological approach.

In Fig.~\ref{kinetic}, R-R separation is equilibrium (except for a very small region $Ba^2/k_\textrm{B}T \lesssim 0.005$). Below the ``AR instability'' line, AR-AR separation is also possible, but metastable. Below the ``R instability'' line, the initial state is not subject to R instability although R-R separation remains the equilibrium state. The red and blue colours show the difference in growth rates, $\Delta \omega$, between R and AR instability modes. For example, red signifies a faster-growing AR instability mode, so that AR-AR separation dominates initial demixing.

For physical parameter ranges \cite{Williamson2014}, neither the R or AR mode is trivially dominant, so that moderate changes to lipid tail length mismatch (affecting the effective value of $J$) or tail structure mismatch (affecting $B$) can determine whether the initial instability after a quench leads to locally symmetric (R) or locally asymmetric (AR) domains. In addition, the long-lived AR-AR states simulated in \cite{Perlmutter2011, Reigada2015} provide \textit{prima facie} evidence that metastable trapping due to failure to nucleate R domains is possible for physical phospholipids. However, due to small simulation sizes in such studies, this could constitute a \textit{stable}, not metastably trapped, state, an issue which we address in the Discussion.

\colorred

Finally we note that, contrary to \cite{Galimzyanov2015}, AR-AR coexistence does \textit{not} require an exactly equimolar (or equal area fractions) mixture in each leaflet. Firstly, an overall composition away from $\Phi^\textrm{t}\!=\!\Phi^\textrm{b}\!=\!0.5$ can lie on one of a set of AR-AR tie-lines running parallel to the central one depicted in Fig.~\ref{schematic} \cite{Williamson2015}. Secondly, even if the overall composition is outside any AR-AR tie-line, AR-AR-R coexistence can occur, in which the presence of some R phase is forced by the composition being far from equimolar, but the R phase coexists with two AR phases \cite{Williamson2015}. In that situation, every region of the bilayer except the R phase is, as for AR-AR, locally asymmetric. Like AR-AR, for typical parameters, AR-AR-R is a metastable state, which could become trapped for strong hydrophobic mismatch or stabilised if domain size is limited (e.g., by simulation size).

\color{black}

\section{Kinetic phase diagram}

Fig.~\ref{kinetic} shows the results of simulations of phase-transition kinetics for varying indirect coupling $J$ and direct coupling $B$, to model varying lipid tail length mismatch and structural mismatch.
The overlaid greyscale dots signify whether registered or antiregistered domains dominate at the end of the simulated time ($t = 10^6$ Monte Carlo Steps) by measuring the degree of registration $\lambda$, with $\lambda \approx 0$ (black) corresponding to AR-AR coexistence and $\lambda \approx 1$ (white) to R-R. Each dot on Fig.~\ref{kinetic} is an average of four independent trajectories.

\begin{figure}[floatfix]
\includegraphics[width=8cm]{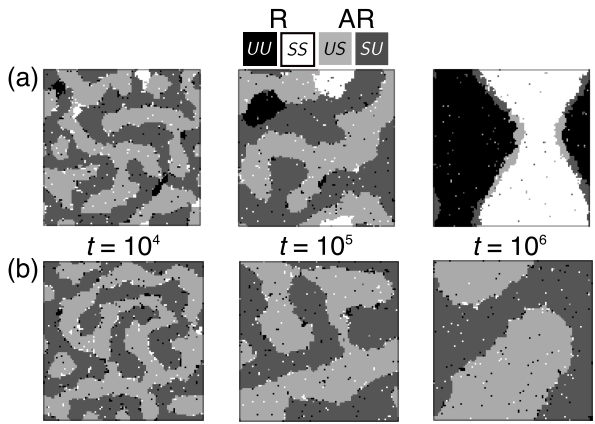}
\caption{\label{nucleation}(a)~Snapshots of a trajectory in which two registered nuclei grow to reach equilibrium. (b)~Trajectory with the same parameters but a different random initial configuration, in which no nuclei survive. Parameters:\ $\Delta_0 = 2\,a$, $\kappa = 3\,a^{-2}k_\textrm{B}T$, $V=0.9\,k_\textrm{B}T$, $J = 4\,a^{-2}k_\textrm{B}T$, $B=0.42\,a^{-2}k_\textrm{B}T$, $\mathcal{L}=100$.}
\end{figure}

This simulated ``kinetic phase diagram'' agrees semi-quantitively with the theoretical linear stability analysis of initial demixing (red/blue on Fig.~\ref{kinetic}):\ inside the blue region, where R-R coexistence should be accessed directly, the simulation exhibits full registration. 
If instead the AR mode is fastest (red), we find two possibilities -- the bilayer may reach R-R coexistence by nucleation out of the AR-AR state, or remain metastably trapped in AR-AR. Fig.~\ref{nucleation}a shows the successful formation of registered nuclei, which grow from the boundaries of antiregistered domains. In Fig.~\ref{nucleation}b, a different random quench in the simulation leads to failure to nucleate despite unchanged parameters, illustrating the stochastic nature of the nucleation process.

Having identified the expected three classes of kinetics, we next model the energetics of nucleating registered domains, to clarify the fates of the metastable AR-AR state (the second and third kinetic classes identified in Fig.~\ref{kinetic}).

\section{Nucleation theory} \label{sec:nuc}

The metastability of AR-AR implies that arbitrarily small composition fluctuations decay. To reach R-R, registered domains must be formed by composition fluctuations sufficient to generate a nucleus of a registered phase large enough for the area-dependent payoff in bulk free energy to outweigh the penalty for hydrophobic mismatch at the edge. We now study the energetics of this nucleation. For simplicity we consider the R nucleus to be compositionally uniform (thus ignoring mixing effects at the domain boundary), and use the response of the thickness profile to mismatch at the nucleus' boundary to calculate the nucleation energetics.

Fig.~\ref{nucleation}a exhibits typical nucleation of registered domains. Our goal is to model the role of hydrophobic mismatch around the edge of R nuclei in determining whether they successfully grow (Fig.~\ref{nucleation}a) or decay (Fig.~\ref{nucleation}b). For the reference (zero free energy) state, we take a registered bilayer at equilibrium with no transmidplane mismatch and domains coarse enough for edges to make a vanishing energy contribution. We then assume that a dominant AR instability mode has led to an initially antiregistered state, which incurs everywhere an energy cost, relative to the registered reference state, due to the direct coupling $B$. We then introduce a circular R nucleus of radius $R$, thus removing the direct coupling energy within the domain's area but introducing a thickness mismatch around its perimeter (Fig.~\ref{leaflets}).

In the continuum limit of the lattice model, the free energy is 
\begin{multline}
G_\textrm{cont} = \frac{1}{2} \int_0^\infty \frac{2 \pi r dr}{a^2} \Bigg( {\kappa}\big(\ell^{\textrm{t}} (r) - \ell_0^{\textrm{t}}(r)\big)^2 \\ + {\kappa}\big(\ell^{\textrm{b}} (r) - \ell_0^{\textrm{b}}(r)\big)^2
 + \tilde{J} a^2 \left( \frac{d d(r)}{dr} \right)^2  + B \big(\ell^\textrm{t}(r) - \ell^\textrm{b}(r)\big)^2 \Bigg)~,
\label{eqn:continuumfunc}
\end{multline}
\noindent where $a$ is the lattice spacing and the $\tilde{J}$ term captures the thickness gradient penalty from the corresponding neighbour interaction (Eq.~\ref{eqn:fun4}) in the limit of small lattice spacing. We neglect contributions from the Ising-like interaction $V$.
It is thought that, as hydrophobic mismatch increases, it becomes the dominant contribution to the line tension of (registered) domains \cite{Garcia2007, Kuzmin2005} (this does not imply that thickness mismatch uniquely determines phase-transition temperature \cite{Bleecker2015}). However, in Appendix~\ref{app:V} we briefly discuss an upper bound for the influence of $V$ on the following nucleation energetics.

The leaflets' compositions are reflected in their ideal thicknesses $\ell_0^{\textrm{t,b}} (r)$. The actual thickness profiles $\ell^{\textrm{t,b}} (r)$ adopted by the leaflets minimise the free energy $G_\textrm{cont}$. The actual thickness difference and total thickness profiles are denoted $\Delta (r) \equiv \ell^{\textrm{t}}(r) - \ell^{\textrm{b}}(r)$ and $d(r) \equiv\ell^{\textrm{t}}(r) + \ell^{\textrm{b}}(r)$. The calculations underlying the following results are detailed in Appendix \ref{app:calculations}.

\subsection{Antiregistered background}

The nucleus appears within a phase-separated AR-AR background. In practice, R domains typically form at AR-AR domain boundaries (Fig.~\ref{nucleation}). The effects of this on nucleation are briefly discussed in Appendix~\ref{app:V}, but for the following calculations we are ignoring the Ising contribution $V$ (which provides the only source of line tension at AR-AR boundaries). Hence, the pattern and orientation of the initial AR domains is unimportant, and we assume a spatially uniform state with $S$ lipids in the top leaflet, $U$ in the bottom.
For simplicity we assume strong compositional segregation so that the phases are approximately pure, with ideal leaflet thicknesses given by
\begin{subequations}
\begin{align}
\ell_0^{\textrm{t, AR}} &= \ell_{S0}~,  \\
 \ell_0^{\textrm{b, AR}} &= \ell_{U0}~.
\end{align}
\label{eqn:before2}
\end{subequations}
Minimisation of Eq.~\ref{eqn:continuumfunc} yields
\begin{equation}
\Delta ^\textrm{AR} = \frac{{\kappa} \Delta_{0}}{{\kappa} + 2B} ~,
\label{eqn:diffbefore}
\end{equation}
\noindent for the difference in actual leaflet thicknesses. The species in each leaflet adapt their tail lengths (hence degree of tail ordering \cite{Komura2004}) to one another's presence, balancing tail stretching with the direct coupling energy. Relaxing the assumption of strong compositional segregation would not qualitatively affect the physics. The physical content of Eq.~\ref{eqn:diffbefore} is simply that transbilayer-mismatched lipids in an AR phase cause a free-energy cost;\ in strong compositional segregation this energy density is (see Eq.~\ref{eqn:delEin})
\begin{equation}
\gamma  = \frac{\Delta_0^2 \kappa B}{2 a^2 (\kappa+2B)}~.
\label{eqn:mismatchenergy}
\end{equation}
Since no mismatches in total ideal thickness exist, $d^\textrm{AR}$ is uniform:\
\begin{equation}
d^\textrm{AR} =  \ell_{S0} + \ell_{U0} = d_0~.
\label{eqn:continuumsolnsbefore}
\end{equation}

\subsection{Introducing a registered nucleus}

\begin{figure}[floatfix]
\includegraphics[width=8.0cm]{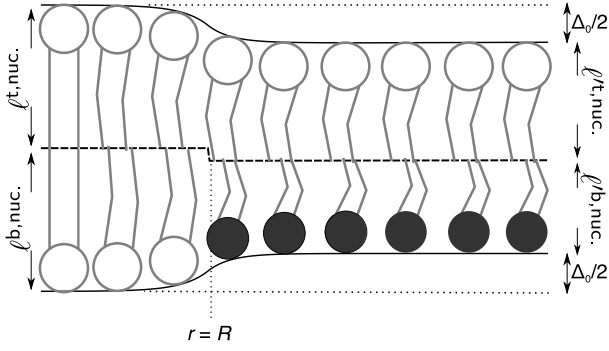}
\caption{\label{leaflets}Schematic structure of a registered $SS$ nucleus introduced to an antiregistered background. The leaflet thicknesses inside and outside the nucleus are calculated from Eq.~\ref{eqn:leaflets}. The midplane (dashed) shifts discontinuously at $r=R$ to maintain smooth outer contours (cf.\ \cite{Galimzyanov2015, Perlmutter2011}). $S$ (light) and $U$ (dark) lipids illustrating composition are superimposed. Primes indicate that leaflet thicknesses outside the nucleus differ from those inside.}
\end{figure}

\noindent We consider the energy change induced by a registered nucleus of the $S$ species (Fig.~\ref{leaflets}), with larger ideal thickness than its surroundings (the energetics are identical for a nucleus of the shorter species $U$):\
\begin{subequations}
\begin{align}
\ell_0^{\textrm{t, nuc.}} (r)  &=  \ell_{S0}~,\\
\ell_0^{\textrm{b, nuc.}} (r) &=
\begin{cases}
\ell_{S0} & \text{if } r\leq R~, \\[5pt]
\ell_{U0} & \text{if } r > R~.\\
\end{cases}
\end{align}
\label{eqn:after2}
\end{subequations}

The top and bottom leaflets now have the same composition (ideal thickness) within the nucleus, removing the inter-leaflet mismatch within the region $r \leq R$ and leading to an area-dependent free energy payoff via the $B$ term of Eq.~\ref{eqn:continuumfunc}. The actual thickness difference profile is now
\begin{equation}
\Delta^\textrm{nuc.}(r) =
\begin{cases}
0 & \text{if } r\leq R~, \\[5pt]
\displaystyle\frac{{\kappa} \Delta _{0}}{{\kappa} + 2B} & \text{if } r > R~.\\
\end{cases}
\label{eqn:diffafter}
\end{equation}
\noindent The discontinuity arises from the discontinuous composition at $r = R$ and the fact that, in contrast to the total thickness $d(r)$, no energy in the Hamiltonian penalises variation in the thickness \textit{difference} $\Delta(r)$. 

If the composition interface were not sharp one would consider the coupling between composition and thickness gradients. Moreover, one could also allow for sliding into antiregistration at a registered domain boundary, which smears out thickness mismatch \cite{Williamson2015, Perlmutter2011, Galimzyanov2015}.
These would not qualitatively alter the fact that the registered nucleus' boundary experiences an energy cost from thickness mismatch, but could help reduce it. 

\begin{figure}[floatfix]
\includegraphics[width=8.0cm]{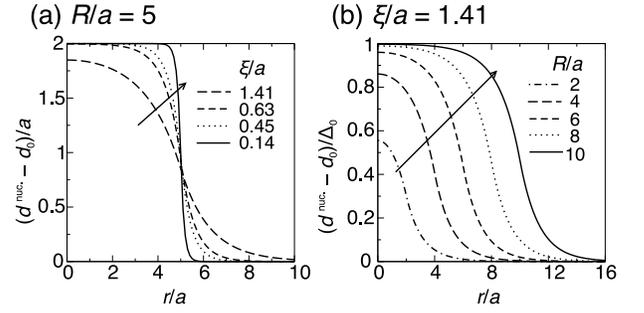}
\caption{\label{ltots}(a)~Calculated profile of total thickness $d^\textrm{nuc.}(r)$ relative to the AR background, when the nucleus is introduced. The \textit{ideal} thickness relative to the surroundings is $\Delta _{0} = 2\,a$. Arrow:\ decreasing decay length $\xi$. (b)~$d^\textrm{nuc.}(r)$ relative to the AR background and normalised by its ideal value inside the nucleus, $\Delta _{0}$. Arrow:\ increasing nucleus size.
}
\end{figure}

Calculating the total thickness profile after the nucleus is introduced yields
\begin{multline}
d^\textrm{nuc.}(r) =\\ 
d_0 + \begin{cases} 
\Delta_0\left[1 - \tfrac{R}{\xi}K_1 \left(\tfrac{R}{\xi} \right) I_0\left(\tfrac{r}{\xi} \right)    \right] & \text{if } r \leq R~,  \\[5pt]
\Delta_0\left[\tfrac{R}{\xi}I_1 \left(\tfrac{R}{\xi} \right) K_0\left(\tfrac{r}{\xi} \right)    \right] & \text{if } r > R~.\\ 
\end{cases}
\label{eqn:continuumsolns}
\end{multline}
\noindent Here, $I_n$ and $K_n$ are $n$th order modified Bessel functions of the first and second kind respectively. Their spatial dependence is controlled by a decay length 
\begin{equation}
\xi \equiv \sqrt{2\tilde{J} a^2/\kappa}~,
\end{equation}
\noindent which quantifies the competition between hydrophobic mismatch and stretching. Eq.~\ref{eqn:continuumsolns} shows how the mismatch in ideal total thickness at $r = R$ distorts the thickness profile both inside and outside the registered nucleus. The decay length $\xi$ controls the lateral distance over which $d^\textrm{nuc.}(r)$ is perturbed (Fig.~\ref{ltots}a). 

For large $\xi$, the R nucleus does not reach its ideal thickness even at its centre, $r/a = 0$. This is an important point;\ for small nuclei $R \sim \xi$, the \textit{effective} thickness mismatch between the nucleus and its surroundings is less than its value $\Delta _{0}$ in the limit of large domain size (Fig.~\ref{ltots}b). The thickness at the centre of the nucleus is given by
\begin{equation}
d^\textrm{nuc.}(0) = d_0 + \Delta_0 \left[1 - \tfrac{R}{\xi} K_1\left(\tfrac{R}{\xi}\right) \right]~.
\label{eqn:maxheight}
\end{equation}
\noindent Hence, small domains have smaller hydrophobic mismatch so experience a smaller effective line tension. This dependence of domain height on domain size could be observed by atomic force microscopy. Since $\tilde{J}$ and ${\kappa}$ will typically be of the same order of magnitude \cite{Williamson2014} such that $\xi \sim a$, it may only be significant for very small domains. However, increasing thickness mismatch of registered domains as they grow has been reported in molecular simulation \cite{Jefferys2014}, as predicted here.

After introducing the nucleus, the individual thicknesses $\ell^\textrm{t, nuc.}(r)$ and $\ell^\textrm{b, nuc.}(r)$ of the top and bottom leaflets are given by
\begin{subequations}
\begin{align}
\ell^\textrm{t, nuc.}(r) &= \displaystyle\tfrac{1}{2}\left(d^\textrm{nuc.}(r) + \Delta ^\textrm{nuc.}(r)\right)~,\\
\ell^\textrm{b, nuc.}(r) &= \displaystyle\tfrac{1}{2}\left(d^\textrm{nuc.}(r) - \Delta ^\textrm{nuc.}(r)\right)~.
\end{align}
\label{eqn:leaflets}
\end{subequations}
\noindent Assuming no empty space inside the bilayer, the leaflet thicknesses fully determine the outer contours of the bilayer once the midplane position is specified. Noting that $\Delta ^\textrm{nuc.}(r)$ (thus $\ell^\textrm{t, nuc.}(r)$ and $\ell^\textrm{b, nuc.}(r)$) is discontinuous at the nucleus boundary $r=R$, we expect this discontinuity to occur in the midplane position to maintain smooth outer contours and minimise hydrophobic exposure \cite{Galimzyanov2015}. This can also be seen in molecular simulations containing AR domains \cite{Perlmutter2011}. The resultant bilayer structure is shown in Fig.~\ref{leaflets}. Outside the R nucleus, the leaflets adapt their thicknesses (thus tail structure) to one another's presence (Eq.~\ref{eqn:diffafter}), i.e.\ the $S$ species are shorter than $\ell_{S0}$ and the $U$ longer than $\ell_{U0}$. Inside the nucleus, the registered region is able to achieve its ideal thickness only once the nucleus grows large enough. 

\begin{figure}[floatfix]
\includegraphics[width=8.0cm]{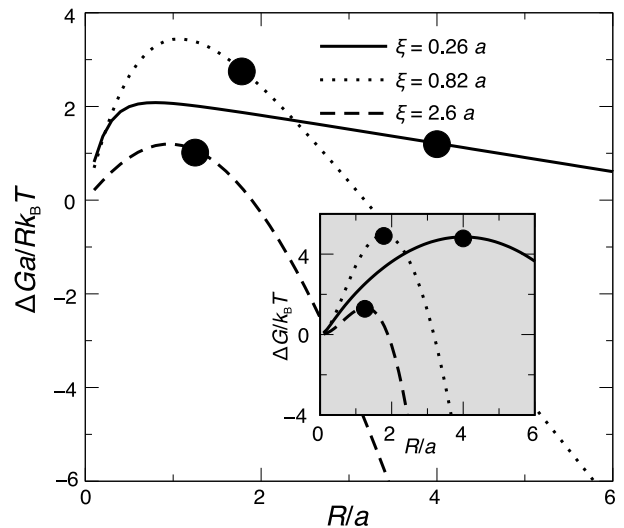}
\caption{\label{curves}Energy $\Delta G$ for a registered nucleus of radius $R$, calculated from Eq.~\ref{eqn:delEexact}, with $\Delta _{0} = 2\,a$ and ${\kappa} = 3\,a^{-2}k_\textrm{B}T$. We vary the decay length $\xi$ by varying $\tilde{J}$ but hold the ratio of indirect and direct couplings constant at $\tilde{J}  / B = 2$ ($J/B = 8$ line in Fig.~\ref{nucbar}). In 2D classical nucleation theory the ratio $\Delta G /R$ would yield a straight line of negative slope. Circles mark critical radii $R^\textrm{*}$. Inset:\ behaviour of $\Delta G$ without normalisation by nucleus size.}
\end{figure}

\subsection{Energy of a registered nucleus}

Given the thickness profiles before and after the nucleus is introduced, the energy required to create a registered nucleus of size $R$ is calculated as (see Appendix~\ref{app:calculations})
\begin{equation}
\Delta G  =  \frac{\Delta _{0}^2{\kappa}}{2a^2} \pi R^2 \left(- \frac{B}{{\kappa}+2B} + I_1\left(\tfrac{R}{\xi}\right)K_1\left(\tfrac{R}{\xi}\right) \right)~.
\label{eqn:delEexact}
\end{equation}
\noindent The negative term $\propto R^2$ arises from the removal of inter-leaflet tail structure mismatch over the domain's area, and the positive term arises from hydrophobic mismatch. For large nucleus size $R$ relative to the decay length $\xi$, the approximations $I_n(x) \sim \exp(x)/\sqrt{2\pi x}$ and $K_n(x) \sim \exp(-x) \sqrt{\pi /2x}$ for large $x$ give $I_1(\tfrac{R}{\xi})K_1(\tfrac{R}{\xi}) \sim \xi/2R$. The hydrophobic mismatch term then becomes overall linear in $R$, acting like a standard line tension $\Gamma = \Delta_0^2\kappa\xi/8a^2$. 
\colorred Hence, for large nuclei or large stiffness $\kappa$, $\Delta G$ behaves as in 2D classical nucleation theory (CNT). For $a\!\approx\!0.8\,\textrm{nm}$, $\kappa\!\approx\!3\,a^{-2}k_\textrm{B}T$, $\xi\!\approx\!a$ and $\Delta_0\!\approx\! a$ at $T\!=\!300\,\textrm{K}$, we estimate $\Gamma \approx 2\,\textrm{pN}$ between R and AR domains or $\Gamma \approx 8\,\textrm{pN}$ between R-R domains of different species, quite close to existing estimates for phospholipids \footnote{Eq.~\ref{eqn:delEexact} refers to a boundary between R-AR domains, whose thickness mismatch is equal to $\Delta_0$, and for large $R$ the estimated parameters give line tension $\Gamma \approx 2\,\textrm{pN}$. For an R-R boundary the thickness mismatch would double to $2 \Delta_0 \!\approx\! 2\,a \!\approx\! 1.6\,\textrm{nm}$, yielding line tension $\Gamma \approx 8\,\textrm{pN}$. For this thickness mismatch, the estimate made in \cite{Garcia2007} is $\Gamma = 6 \pm 2\,\textrm{pN}$, using existing theory \cite{Kuzmin2005} which has been shown to yield satisfactory agreement with available experiments}.
Thus, although our simplified model does not capture all details of hydrophobic mismatch and the concomitant bilayer deformation \cite{Galimzyanov2015}, the associated energy scale is well captured. 

\color{black}
Some illustrative nucleation energy curves are shown in Fig.~\ref{curves}. For CNT in 2D, the ratio $\Delta G / R$ would be a straight line of negative slope. Deviation from a straight line indicates deviation from CNT. If $\xi$ is small (weak hydrophobic mismatch compared to stiffness), the behaviour is CNT-like over most of $R$;\ moreover the critical radius $R^\textrm{*}$ (defined as the value where $\Delta G(R)$ is maximised) occurs deep into the CNT-like regime. For larger $\xi$ (stronger hydrophobic length mismatch relative to stiffness) the critical radius no longer occurs in the CNT-like regime, so non-CNT effects arising from nonlocal thickness deformation influence nucleation. 

\begin{figure}[floatfix]
\includegraphics[width=8.0cm]{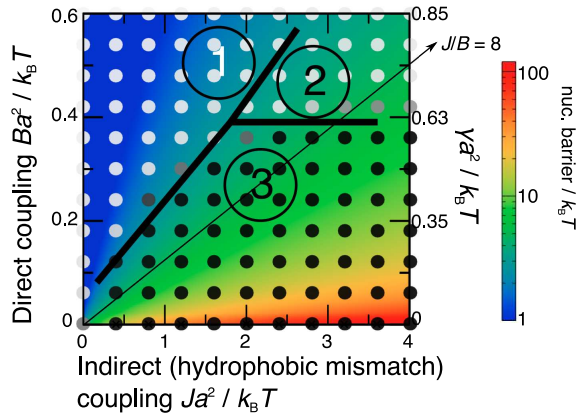}
\caption{\label{nucbar}Simulated kinetic phase diagram (notation as in Fig.~\ref{kinetic}), overlaid on the theoretical nucleation barrier for an R domain nucleating from an AR state ($\Delta_0 = 2\,a$, ${\kappa} = 3\,a^{-2}k_\textrm{B}T$). The arrowed line shows $J/B = 8$, as used in Fig.~\ref{curves}.}
\end{figure}

\begin{figure}[floatfix]
\includegraphics[width=8.0cm]{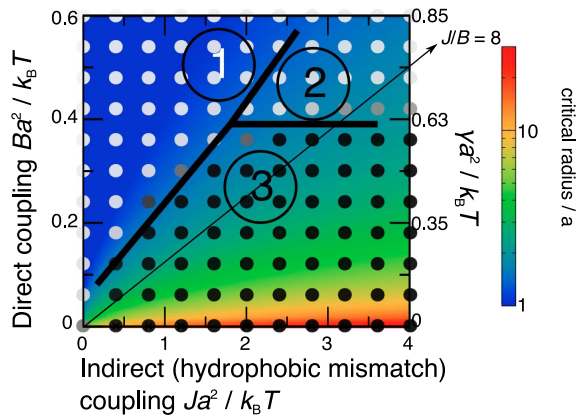}
\caption{\label{critrad}As Fig.~\ref{nucbar} but showing the calculated critical radius for R domain nucleation.}
\end{figure}

\subsection{Nucleation barrier and critical radius landscapes} \label{sec:landscapes}

Fig.~\ref{curves} (inset) implies that increasing $J$ for a fixed ratio $J / B$ and fixed $\kappa$ (equivalently, increasing $\xi$) first increases, then reduces, the nucleation barrier. In the limit of large $\kappa$ one would expect increasing $J$ and $B$ to simply increase the energy scale, hence the nucleation barrier. But when $J$ is comparable to $\kappa$, the mechanism illustrated in Fig.~\ref{ltots}b takes over;\ increasing $\xi \equiv \sqrt{2\tilde{J} a^2/\kappa}$ reduces the effective thickness mismatch for small nuclei, thus reducing the hydrophobic penalty for nucleation and reducing the nucleation barrier.

The nucleation theory can be related to the simulations by plotting landscapes of the nucleation barrier $\Delta G(R^{*})$ (Fig.~\ref{nucbar}) and critical radius $R^{*}$ (Fig.~\ref{critrad}). A small barrier clearly facilitates nucleation, although the role of the critical radius is more subtle \cite{Ryu2010, Vehkamaki2006, Zeldovich2007}. One important factor is that for smaller critical radii, regions randomly enriched in registration from the initial quench are more likely to be near or above the critical radius, so can grow more easily. 

The predicted energetics for nucleation are consistent with the classes of kinetics identified in Fig.~\ref{kinetic}. Where the R instability mode is fastest, nucleation energetics are irrelevant since the equilibrium phases are formed immediately. If the AR mode is fastest, a successful subsequent transition to equilibrium R-R coexistence generally occurs where the predicted nucleation barrier and critical radius are smaller. \colorred 
As we have used a relatively large value of the mismatch parameter $\Delta_0 = 2\,a$, we briefly mention the dependence on it \footnote{Simulations with smaller values of the mismatch parameter $\Delta_0$ have been found to yield the same three kinetic classes as studied here \cite{Williamson2015}.}. The critical radii are independent of $\Delta_0$, because $\Delta_0$ represents \textit{both} tail length and structure mismatch so affects both the indirect and direct coupling strengths. Thus the only change to Fig.~\ref{critrad} would be in the effective value of $\gamma$ shown on the secondary axis, which scales as $\Delta_0^2$ (Eq.~\ref{eqn:mismatchenergy}). On Fig.~\ref{nucbar}, in addition to the change in effective $\gamma$, the nucleation barrier $\Delta G(R^{*})$ scales as $\Delta_0^2$. 
\color{black}

\section{Discussion}

In this paper we have studied phase-separation kinetics in a bilayer subject to competing symmetric and asymmetric instabilities. We have modelled the nucleation of symmetric (registered, R) phases where the initial spinodal instability has led to antiregistered AR-AR coexistence (local asymmetry everywhere). We made use of a microscopic lattice model that allows study of both phase equilibria and phase-transition kinetics, through coarse-grained theory and direct simulation. 
For realistic parameters, the competing symmetric and asymmetric instabilities can be comparable in strength, so that changes to molecule properties can tip the balance \cite{Williamson2014}. 

If the fastest-growing instability mode is R, spinodal decomposition directly into equilibrium R-R coexistence takes place (class 1 Fig.~\ref{kinetic}). If the AR mode grows fastest then a metastable AR-AR state forms, and nucleation is required to reach equilibrium (class 2 Fig.~\ref{kinetic}). 
Unsuccessful nucleation leaves the bilayer metastably trapped (class 3 Fig.~\ref{kinetic}). 

The key parameters for the initial demixing and subsequent nucleation energetics are the indirect inter-leaflet coupling $J$ via hydrophobic mismatch, favouring antiregistration, and the direct coupling $B$ which favours registration. Physically, the hydrophobic tail length mismatch or structure mismatch would affect the effective values of $J$ or $B$ respectively.
Both $J$ and $B$ couple to the stiffness $\kappa$, which determines how easily mixed lipids can change their tail length and structure to adapt to one another's presence. Here we have focused on an equimolar mixture in both leaflets such that AR-AR coexistence competes with R-R;\ the kinetic considerations are qualitatively similar for other overall compositions subject to competing instability modes, in which other states including three-phase coexistence (AR-AR-R or R-R-AR) can enter \cite{Williamson2015}.

Nucleation implies a critical radius, which a registered domain must exceed so that the penalty of thickness mismatch at the perimeter does not outweigh the bulk free energy gain. This does not mean domains automatically become registered beyond a certain size, as is sometimes implied \cite{May2009, Wallace2006, Reigada2015}. \colorred 
If nucleation is energetically prohibitive, the mere presence of large coarsening antiregistered domains does not guarantee registered domains will ever form, just as assembling a large volume of supercooled water does not guarantee a supercritical ice nucleus will form. That said, it is intriguing to consider whether some hydrodynamic or curvature-mediated mechanism (neither of which are included here) could dynamically assist very large antiregistered domains to overcome the nucleation barrier and become registered.

In the present simulations it appears that a nucleation barrier $\Delta G(R^{*}) \gtrsim 5\,k_\textrm{B}T$ inhibits nucleation on the simulated timescale for this simulation size. In principle the nucleation rate $I$ (per unit area per unit time) is determined by $\Delta G(R^{*})$ via $I = I_0 \exp{(-\Delta G(R^{*})/k_\textrm{B}T)}$, although the unknown kinetic prefactor $I_0$ severely limits the quantitative utility of such a picture. We also expect the critical radius $R^{*}$ to determine how large \textit{random} regions of inter-leaflet symmetry during the initial quench must be in order for them to grow. The critical radius also captures a size-dependence of potential relevance to cell membrane rafts and clusters, discussed below.

\color{black}
By examining the effect of thickness mismatch at the boundary of a nucleating registered domain, we showed that the process departs strongly from classical nucleation theory due to deformation of the thickness profile, which reduces the effective line tension of small nuclei. For small nuclei, the deformation essentially spreads over the whole nucleus, reducing the effective hydrophobic mismatch. To our knowledge this behaviour has not been noted in previous theories which tended to focus on domains large enough to treat the boundary as a straight line \cite{Kuzmin2005, Galimzyanov2015}. 
The prediction is supported by a recent molecular simulation of coarsening of a quenched bilayer \cite{Jefferys2014}, where registered liquid-ordered and liquid-disordered domains became respectively thicker and thinner as they grew through time. 

\colorred 
The simulation method does not include hydrodynamics, which are expected to dominate domain coarsening beyond a lengthscale $\sim 10^{-6}\, \textrm{m}$ \cite{Fan2010} such that the purely diffusive dynamics simulated here would no longer apply.
However, this lengthscale is far beyond both the size simulated here and the predicted critical radii (Fig.~\ref{critrad}). Note that hydrodynamics cannot influence the free-energy landscape of the system, and thus cannot change the competing metastable and equilibrium states.
\color{black}

The phase equilibria of ``macroscopic'' domains (which coarsen until limited by bilayer size) are not influenced by edge energies and, for most reasonable parameters, we predict that R phases are lower in free energy so that R-R is the equilibrium state in bulk \cite{Williamson2014}, as seen in fluorescence studies of large domains \cite{Korlach1999, Dietrich2001}. At the opposite end of the size spectrum, experiments reveal pairwise antiregistration at the single-lipid level \cite{Zhang2004, Zhang2007}, as also reported in simulation \cite{Stevens2005} and predicted by our mean-field theory \cite{Williamson2014}. In between these regimes, edge energies can influence the equilibrium state;\ for example, the AR-AR coexistence in \cite{Perlmutter2011, Reigada2015} is probably metastable in the limit of large size but could be stabilised if the simulation box is too small to accommodate a supercritical registered domain.

This has important physical consequences, because cell membrane rafts or clusters are not macroscopic. Small domain size \textit{in vivo} could be due to elastic repulsion \cite{Meinhardt2013}, hybrid lipids \cite{Palmieri2014}, critical fluctuations \cite{Veatch2007}, active recycling \cite{Fan2008, Turner2005} or another mechanism. In either case, it is crucial to recognise that edge energies could influence the thermodynamic preference and even stabilise an AR-AR type of domain formation, in which one leaflet's local enrichment in longer species (relative to the background in that leaflet) colocalises with a relative depletion of such species in the other leaflet.

Although the cell membrane is maintained out of equilibrium, thermodynamic driving forces can be expected to play a role. The potential importance of this size-dependence is underlined by the fact that the estimated critical radii (Fig.~\ref{critrad}) can be of the order of putative lipid raft sizes \cite{Lingwood2010}. 
The basic biophysical question is:\ if a cluster of longer lipids and proteins exists in one leaflet, does it colocalise a similar cluster in the opposite leaflet to maintain transbilayer structural similarity, or does it choose shorter lipids and proteins to maintain uniform thickness? Our work implies that finite-size effects, metastable states and phase-transition kinetics can be key in determining the answer.

\appendix

\section{\label{app:calculations}Calculations for nucleation theory}
In the nucleation theory the response of the actual profiles $\ell^{\textrm{t,b}} (r)$ to variations in the ideal thicknesses $\ell_0^{\textrm{t,b}} (r)$ is calculated as follows. Varying $G_\textrm{cont}$ (Eq.~\ref{eqn:continuumfunc}) with respect to $\ell^{\textrm{t}}$ and $\ell^{\textrm{b}}$ gives a pair of differential equations that can be combined to yield independent equations for the thickness difference $\Delta (r) \equiv \ell^{\textrm{t}}(r) - \ell^{\textrm{b}}(r)$ and the total $d(r)$:
\begin{subequations}
\begin{align}
{\kappa} \left(\Delta (r) - \Delta _{0}^*(r)\right) + 2B\Delta (r) &= 0~,
\label{eqn:odedelta}\\
r {\kappa} \left(d_0^* (r) - d(r) \right) + 2 \tilde{J} a^2 \frac{d}{dr} \left( r \frac{d d(r)}{dr} \right) &= 0~.
\label{eqn:odetotal}
\end{align}
\end{subequations}
\noindent In the above we have used the ideal thickness difference and total profiles, $\Delta _{0}^*(r) \equiv  \ell_0^{\textrm{t}} (r) -\ell_0^{\textrm{b}} (r)$ and $d_0^* (r) \equiv  \ell_0^{\textrm{t}} (r) +\ell_0^{\textrm{b}} (r)$.

In the AR background the $S$ and $U$ species' lengths are not equal to their ideal values -- the direct coupling $B$ encourages equality of tail length (thus degree of ordering) across the bilayer. Eq.~\ref{eqn:odedelta} is solved by
\begin{equation}
\Delta (r) = \frac{{\kappa} \Delta _{0}^* (r)}{{\kappa} + 2B}~.
\end{equation}
\noindent Inserting Eq.~\ref{eqn:before2} for the ideal thicknesses yields Eq.~\ref{eqn:diffbefore}. The total thickness before the nucleus is introduced is uniform (Eq.~\ref{eqn:continuumsolnsbefore}).

After introducing the R nucleus, the difference in ideal leaflet thicknesses is zero within the R nucleus ($\Delta _0^*(r) = 0$ for $r \leq R$), which gives Eq.~\ref{eqn:diffafter}. The total thickness profile is found by solving Eq.~\ref{eqn:odetotal} (which is expressible as a modified Bessel's equation of order zero). We require that the gradient vanishes at the centre of the nucleus ($ d \,d(r)/ dr \rvert_{r = 0} = 0$), and that the total thickness approaches its ideal value away from the domain ($d(\infty) = \ell_{S0} + \ell_{U0} = d_0$). The profile $d(r)$ and its gradient are required to be continuous at $r = R$ in order to match the profiles inside and outside the nucleus. With ideal thicknesses given by Eq.~\ref{eqn:after2}, this yields Eq.~\ref{eqn:continuumsolns}.

To calculate the energy $\Delta G$ required to introduce the registered nucleus, we consider the initial energy $G^\textrm{AR}_\textrm{cont}$ from inserting Eqs.~\ref{eqn:diffbefore} and \ref{eqn:continuumsolnsbefore} into Eq.~\ref{eqn:continuumfunc}, and the energy $G^\textrm{nuc.}_\textrm{cont}$ when the nucleus is introduced, from inserting Eqs.~\ref{eqn:diffafter} and \ref{eqn:continuumsolns} instead. The energy for introducing the nucleus is then 
\begin{equation}
\Delta G = G^\textrm{nuc.}_\textrm{cont} - G^\textrm{AR}_\textrm{cont}~.
\label{eqn:delE}
\end{equation}
\noindent 
Since Eqs.~\ref{eqn:diffafter} and \ref{eqn:continuumsolns} split at the nucleus boundary $r=R$, we evaluate separate contributions to $\Delta G$ outside and inside the nucleus and find
\begin{multline}
\Delta G^\textrm{out} =  \int_R^\infty \frac{2 \pi r dr}{a^2} \Bigg(\frac{\tilde{J} a^2}{2}\left(\frac{d}{dr} d^\textrm{nuc.}(r)\right)^2\\
 + \frac{{\kappa}}{4}\left(d^\textrm{nuc.}(r) - d_0\right)^2 \Bigg)~,
\label{eqn:delEout}
\end{multline}
\begin{multline}
\Delta G^\textrm{in} =  \int_0^R \frac{2 \pi r dr}{a^2} \Bigg(\frac{\tilde{J}a^2}{2}\left(\frac{d}{dr} d^\textrm{nuc.}(r)\right)^2\\  + \frac{{\kappa}}{4}\left(d^\textrm{nuc.}(r) - 2\ell_{S0}\right)^2 
- \frac{\Delta _{0}^2 \kappa B}{2 ({\kappa}+2B)}    \Bigg)~.
\label{eqn:delEin}
\end{multline}
\noindent $\Delta G^\textrm{out}$ contains only positive contributions, resulting from deformation of the total thickness profile by the nucleus. $\Delta G^\textrm{in}$ contains similar deformation terms, but also a negative term resulting from the fact that there is now no inter-leaflet mismatch in the region $r \leq R$. The sum $\Delta G^\textrm{in} +\Delta G^\textrm{out}$ leads to Eq.~\ref{eqn:delEexact}. 

Note that the negative term in $\Delta G^\textrm{in}$ yields the estimated inter-leaflet mismatch energy density $\gamma$ under our microscopic definition (Eq.~\ref{eqn:mismatchenergy}) \cite{Williamson2014}. Near the strong-segregation regime this is a close approximation to the difference in bulk free-energy density between registered and antiregistered phases (which can be used to construct an alternative ``macroscopic'' definition of $\gamma$ \cite{Williamson2014}), becoming exact in the strong-segregation limit assumed in Section~\ref{sec:nuc}. However, in principle one could replace this term with the actual free-energy difference in cases where the correspondence between microscopic and macroscopic $\gamma$ breaks down \cite{Williamson2014}.

\section{\label{app:V}Effect of $V$ on nucleation}

\begin{figure}[floatfix]
\includegraphics[width=8cm]{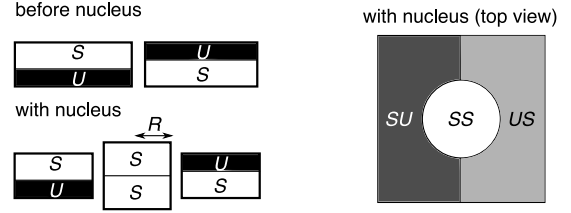}
\caption{\label{appcartoon}Schematic of an $SS$ R nucleus of radius $R$ forming at the boundary between AR-AR domains.}
\end{figure}

\colorred 

In Section~\ref{sec:nuc} we ignored the effect of $V$ (headgroup interactions) on the nucleation energetics. This applies best where hydrophobic mismatch is the dominant source of line tension \cite{Kuzmin2005}. Here we sketch an upper bound on the contribution of $V$ to nucleation energetics of a registered domain. Fig.~\ref{appcartoon} shows an R domain nucleating at a boundary between AR domains (cf.\ Fig.~\ref{nucleation}). This relieves some AR-AR interfacial energy cost (like a particle in a Pickering emulsion).

Assuming the original AR-AR boundary is straight and the R nucleus is a circle, the length of AR-AR boundary removed by an R domain of radius $R$ is equal to $2R$. The length of R-AR boundary \textit{introduced} by the nucleus is $2\pi R$. It is extremely difficult to estimate the line tension caused by $V$, because i)~the separated phases will not be fully pure and ii)~the interface will relax its energy by smearing the composition change over some finite distance, and iii)~this will be \textit{coupled} to the thickness gradients. However, an upper bound can be obtained by assuming, as in the main text, that the phases are strongly-segregated and that the compositional interface is sharp. With these assumptions, the line tension of the AR-AR boundary is $2V/a$, and the contribution from $V$ to an R-AR boundary is $V/a$. Subtracting the AR-AR energy contribution and adding the R-AR contribution, the correction to Eq.~\ref{eqn:delEexact} would be:
\begin{equation}
\Delta G \to \Delta G + \frac{2V}{a} (\pi -2) R~.
\label{eqn:correction}
\end{equation}
\noindent Hence, the contribution of $V$ is always to increase the nucleation energy (thus both the critical radius and nucleation barrier) of registered domains.
For the parameters ranges covered by Fig.~\ref{kinetic} (assuming $V=0.9\,k_\textrm{B}T$ as used in the simulations), we find that including the upper-bound correction from $V$ increases the critical radius $R^{*}$ by at most $\sim 40\,\%$ for $J \gtrsim 2\,a^{-2}k_\textrm{B}T$, from the corresponding value on Fig.~\ref{critrad}. (If $\Delta_0$ is reduced to $1\,a$, reflecting decreased tail hydrophobic and structural mismatch so that $V$ is proportionally more important relative to the indirect and direct couplings, the increase can be $\sim 100\,\%$). For $\Delta G(R^*)$, if we assume that the original $R^*$ is not altered much by including $V$, the associated increase in the nucleation barrier is estimated as $(2V/a) (\pi -2) R^*$. Alternatively, explicitly taking into account the change in $R^*$ resulting from including $V$, the increase in $\Delta G(R^*)$ in the parameter range covered by Fig.~\ref{kinetic} is up to $\sim 100\,\%$  for $J \gtrsim 2\,a^{-2}k_\textrm{B}T$, from the corresponding value in Fig.~\ref{nucbar}.

We reiterate that Eq.~\ref{eqn:correction} must overestimate the contribution $V$ to nucleation energetics, possibly quite severely, so that the argument outlined here provides only a rough upper bound.

\color{black}

\begin{acknowledgments} 
We acknowledge discussions with members of the EPSRC CAPITALS programme grant, and the input of anonymous referees. Lucas Vieira Barbosa is thanked for assistance in plot design.
This work was initiated at the University of Leeds (United Kingdom), and was funded by EPSRC grant EP/J017566/1 and by Georgetown University. PDO gratefully acknowledges the support of the Ives endowment.
\end{acknowledgments} 

\bibliography{bibliography}

\begin{thebibliography}{49}%
\makeatletter
\providecommand \@ifxundefined [1]{%
 \@ifx{#1\undefined}
}%
\providecommand \@ifnum [1]{%
 \ifnum #1\expandafter \@firstoftwo
 \else \expandafter \@secondoftwo
 \fi
}%
\providecommand \@ifx [1]{%
 \ifx #1\expandafter \@firstoftwo
 \else \expandafter \@secondoftwo
 \fi
}%
\providecommand \natexlab [1]{#1}%
\providecommand \enquote  [1]{``#1''}%
\providecommand \bibnamefont  [1]{#1}%
\providecommand \bibfnamefont [1]{#1}%
\providecommand \citenamefont [1]{#1}%
\providecommand \href@noop [0]{\@secondoftwo}%
\providecommand \href [0]{\begingroup \@sanitize@url \@href}%
\providecommand \@href[1]{\@@startlink{#1}\@@href}%
\providecommand \@@href[1]{\endgroup#1\@@endlink}%
\providecommand \@sanitize@url [0]{\catcode `\\12\catcode `\$12\catcode
  `\&12\catcode `\#12\catcode `\^12\catcode `\_12\catcode `\%12\relax}%
\providecommand \@@startlink[1]{}%
\providecommand \@@endlink[0]{}%
\providecommand \url  [0]{\begingroup\@sanitize@url \@url }%
\providecommand \@url [1]{\endgroup\@href {#1}{\urlprefix }}%
\providecommand \urlprefix  [0]{URL }%
\providecommand \Eprint [0]{\href }%
\providecommand \doibase [0]{http://dx.doi.org/}%
\providecommand \selectlanguage [0]{\@gobble}%
\providecommand \bibinfo  [0]{\@secondoftwo}%
\providecommand \bibfield  [0]{\@secondoftwo}%
\providecommand \translation [1]{[#1]}%
\providecommand \BibitemOpen [0]{}%
\providecommand \bibitemStop [0]{}%
\providecommand \bibitemNoStop [0]{.\EOS\space}%
\providecommand \EOS [0]{\spacefactor3000\relax}%
\providecommand \BibitemShut  [1]{\csname bibitem#1\endcsname}%
\let\auto@bib@innerbib\@empty
\bibitem [{\citenamefont {Williamson}\ and\ \citenamefont
  {Olmsted}(2015)}]{Williamson2014}%
  \BibitemOpen
  \bibfield  {author} {\bibinfo {author} {\bibfnamefont {J.~J.}\ \bibnamefont
  {Williamson}}\ and\ \bibinfo {author} {\bibfnamefont {P.~D.}\ \bibnamefont
  {Olmsted}},\ }\href {\doibase 10.1016/j.bpj.2015.03.016} {\bibfield
  {journal} {\bibinfo  {journal} {Biophys. J.}\ }\textbf {\bibinfo {volume}
  {108}},\ \bibinfo {pages} {1963} (\bibinfo {year} {2015})}\BibitemShut
  {NoStop}%
\bibitem [{\citenamefont {{Williamson}}\ and\ \citenamefont
  {{Olmsted}}(2015)}]{Williamson2015}%
  \BibitemOpen
  \bibfield  {author} {\bibinfo {author} {\bibfnamefont {J.~J.}\ \bibnamefont
  {{Williamson}}}\ and\ \bibinfo {author} {\bibfnamefont {P.~D.}\ \bibnamefont
  {{Olmsted}}},\ }\href@noop {} {} (\bibinfo {year} {2015}),\ \bibinfo {note}
  {arXiv:1505.02078},\ \Eprint {http://arxiv.org/abs/1505.02078}
  {arXiv:1505.02078} \BibitemShut {NoStop}%
\bibitem [{\citenamefont {Garb{\`e}s~Putzel}\ and\ \citenamefont
  {Schick}(2008)}]{Putzel2008}%
  \BibitemOpen
  \bibfield  {author} {\bibinfo {author} {\bibfnamefont {G.}~\bibnamefont
  {Garb{\`e}s~Putzel}}\ and\ \bibinfo {author} {\bibfnamefont {M.}~\bibnamefont
  {Schick}},\ }\href
  {http://linkinghub.elsevier.com/retrieve/pii/S0006349508706854} {\bibfield
  {journal} {\bibinfo  {journal} {Biophys. J.}\ }\textbf {\bibinfo {volume}
  {94}},\ \bibinfo {pages} {869} (\bibinfo {year} {2008})}\BibitemShut
  {NoStop}%
\bibitem [{\citenamefont {Garb{\`e}s~Putzel}\ \emph {et~al.}(2011)\citenamefont
  {Garb{\`e}s~Putzel}, \citenamefont {Uline}, \citenamefont {Szleifer},\ and\
  \citenamefont {Schick}}]{Putzel2011}%
  \BibitemOpen
  \bibfield  {author} {\bibinfo {author} {\bibfnamefont {G.}~\bibnamefont
  {Garb{\`e}s~Putzel}}, \bibinfo {author} {\bibfnamefont {M.~J.}\ \bibnamefont
  {Uline}}, \bibinfo {author} {\bibfnamefont {I.}~\bibnamefont {Szleifer}}, \
  and\ \bibinfo {author} {\bibfnamefont {M.}~\bibnamefont {Schick}},\ }\href
  {http://linkinghub.elsevier.com/retrieve/pii/S0006349511000671} {\bibfield
  {journal} {\bibinfo  {journal} {Biophys. J.}\ }\textbf {\bibinfo {volume}
  {100}},\ \bibinfo {pages} {996} (\bibinfo {year} {2011})}\BibitemShut
  {NoStop}%
\bibitem [{\citenamefont {May}(2009)}]{May2009}%
  \BibitemOpen
  \bibfield  {author} {\bibinfo {author} {\bibfnamefont {S.}~\bibnamefont
  {May}},\ }\href {\doibase 10.1039/B901647C} {\bibfield  {journal} {\bibinfo
  {journal} {Soft Matter}\ }\textbf {\bibinfo {volume} {5}},\ \bibinfo {pages}
  {3148} (\bibinfo {year} {2009})}\BibitemShut {NoStop}%
\bibitem [{\citenamefont {Collins}\ and\ \citenamefont
  {Keller}(2008)}]{Collins2008}%
  \BibitemOpen
  \bibfield  {author} {\bibinfo {author} {\bibfnamefont {M.~D.}\ \bibnamefont
  {Collins}}\ and\ \bibinfo {author} {\bibfnamefont {S.~L.}\ \bibnamefont
  {Keller}},\ }\href {\doibase 10.1073/pnas.0702970105} {\bibfield  {journal}
  {\bibinfo  {journal} {Proc. Natl. Acad. Sci.}\ }\textbf {\bibinfo {volume}
  {105}},\ \bibinfo {pages} {124} (\bibinfo {year} {2008})}\BibitemShut
  {NoStop}%
\bibitem [{\citenamefont {Risselada}\ and\ \citenamefont
  {Marrink}(2008)}]{Risselada2008}%
  \BibitemOpen
  \bibfield  {author} {\bibinfo {author} {\bibfnamefont {H.~J.}\ \bibnamefont
  {Risselada}}\ and\ \bibinfo {author} {\bibfnamefont {S.~J.}\ \bibnamefont
  {Marrink}},\ }\href {\doibase 10.1073/pnas.0807527105} {\bibfield  {journal}
  {\bibinfo  {journal} {Proc. Natl. Acad. Sci.}\ }\textbf {\bibinfo {volume}
  {105}},\ \bibinfo {pages} {17367} (\bibinfo {year} {2008})}\BibitemShut
  {NoStop}%
\bibitem [{\citenamefont {Stevens}(2005)}]{Stevens2005}%
  \BibitemOpen
  \bibfield  {author} {\bibinfo {author} {\bibfnamefont {M.~J.}\ \bibnamefont
  {Stevens}},\ }\href {\doibase 10.1021/ja043611q} {\bibfield  {journal}
  {\bibinfo  {journal} {J. Am. Chem. Soc.}\ }\textbf {\bibinfo {volume}
  {127}},\ \bibinfo {pages} {15330} (\bibinfo {year} {2005})}\BibitemShut
  {NoStop}%
\bibitem [{\citenamefont {Perlmutter}\ and\ \citenamefont
  {Sachs}(2011)}]{Perlmutter2011}%
  \BibitemOpen
  \bibfield  {author} {\bibinfo {author} {\bibfnamefont {J.~D.}\ \bibnamefont
  {Perlmutter}}\ and\ \bibinfo {author} {\bibfnamefont {J.~N.}\ \bibnamefont
  {Sachs}},\ }\href {\doibase 10.1021/ja106626r} {\bibfield  {journal}
  {\bibinfo  {journal} {J. Am. Chem. Soc.}\ }\textbf {\bibinfo {volume}
  {133}},\ \bibinfo {pages} {6563} (\bibinfo {year} {2011})}\BibitemShut
  {NoStop}%
\bibitem [{\citenamefont {Reigada}\ and\ \citenamefont
  {Sagu{\'e}s}(2015)}]{Reigada2015}%
  \BibitemOpen
  \bibfield  {author} {\bibinfo {author} {\bibfnamefont {R.}~\bibnamefont
  {Reigada}}\ and\ \bibinfo {author} {\bibfnamefont {F.}~\bibnamefont
  {Sagu{\'e}s}},\ }\href {\doibase 10.1098/rsif.2015.0197} {\bibfield
  {journal} {\bibinfo  {journal} {J. R. Soc. Interface}\ }\textbf {\bibinfo
  {volume} {12}} (\bibinfo {year} {2015}),\ 10.1098/rsif.2015.0197}\BibitemShut
  {NoStop}%
\bibitem [{\citenamefont {Zhang}\ \emph {et~al.}(2004)\citenamefont {Zhang},
  \citenamefont {Jing}, \citenamefont {Tokutake},\ and\ \citenamefont
  {Regen}}]{Zhang2004}%
  \BibitemOpen
  \bibfield  {author} {\bibinfo {author} {\bibfnamefont {J.}~\bibnamefont
  {Zhang}}, \bibinfo {author} {\bibfnamefont {B.}~\bibnamefont {Jing}},
  \bibinfo {author} {\bibfnamefont {N.}~\bibnamefont {Tokutake}}, \ and\
  \bibinfo {author} {\bibfnamefont {S.~L.}\ \bibnamefont {Regen}},\ }\href
  {\doibase 10.1021/ja046892a} {\bibfield  {journal} {\bibinfo  {journal} {J.
  Am. Chem. Soc.}\ }\textbf {\bibinfo {volume} {126}},\ \bibinfo {pages}
  {10856} (\bibinfo {year} {2004})}\BibitemShut {NoStop}%
\bibitem [{\citenamefont {Wagner}\ \emph {et~al.}(2007)\citenamefont {Wagner},
  \citenamefont {Loew},\ and\ \citenamefont {May}}]{Wagner2007}%
  \BibitemOpen
  \bibfield  {author} {\bibinfo {author} {\bibfnamefont {A.~J.}\ \bibnamefont
  {Wagner}}, \bibinfo {author} {\bibfnamefont {S.}~\bibnamefont {Loew}}, \ and\
  \bibinfo {author} {\bibfnamefont {S.}~\bibnamefont {May}},\ }\href
  {http://linkinghub.elsevier.com/retrieve/pii/S000634950771679X} {\bibfield
  {journal} {\bibinfo  {journal} {Biophys. J.}\ }\textbf {\bibinfo {volume}
  {93}},\ \bibinfo {pages} {4268} (\bibinfo {year} {2007})}\BibitemShut
  {NoStop}%
\bibitem [{\citenamefont {Kusumi}\ \emph {et~al.}(2004)\citenamefont {Kusumi},
  \citenamefont {Koyama-Honda},\ and\ \citenamefont {Suzuki}}]{Kusumi2004}%
  \BibitemOpen
  \bibfield  {author} {\bibinfo {author} {\bibfnamefont {A.}~\bibnamefont
  {Kusumi}}, \bibinfo {author} {\bibfnamefont {I.}~\bibnamefont
  {Koyama-Honda}}, \ and\ \bibinfo {author} {\bibfnamefont {K.}~\bibnamefont
  {Suzuki}},\ }\href {\doibase 10.1111/j.1600-0854.2004.0178.x} {\bibfield
  {journal} {\bibinfo  {journal} {Traffic}\ }\textbf {\bibinfo {volume} {5}},\
  \bibinfo {pages} {213} (\bibinfo {year} {2004})}\BibitemShut {NoStop}%
\bibitem [{\citenamefont {Ostwald}(1897)}]{Ostwald}%
  \BibitemOpen
  \bibfield  {author} {\bibinfo {author} {\bibfnamefont {W.}~\bibnamefont
  {Ostwald}},\ }\href@noop {} {\bibfield  {journal} {\bibinfo  {journal} {Z.
  Phys. Chem.}\ }\textbf {\bibinfo {volume} {22}},\ \bibinfo {pages} {289}
  (\bibinfo {year} {1897})}\BibitemShut {NoStop}%
\bibitem [{Note1()}]{Note1}%
  \BibitemOpen
  \bibinfo {note} {In a real system in which lipid species' molecular areas may
  differ, $\Phi ^\protect \textrm {t}\protect \tmspace -\thinmuskip
  {.1667em}=\protect \tmspace -\thinmuskip {.1667em}\Phi ^\protect \textrm
  {b}\protect \tmspace -\thinmuskip {.1667em}=\protect \tmspace -\thinmuskip
  {.1667em}0.5$ refers instead to an equal area fractions mixture in each
  leaflet.}\BibitemShut {Stop}%
\bibitem [{\citenamefont {Komura}\ \emph {et~al.}(2004)\citenamefont {Komura},
  \citenamefont {Shirotori}, \citenamefont {Olmsted},\ and\ \citenamefont
  {Andelman}}]{Komura2004}%
  \BibitemOpen
  \bibfield  {author} {\bibinfo {author} {\bibfnamefont {S.}~\bibnamefont
  {Komura}}, \bibinfo {author} {\bibfnamefont {H.}~\bibnamefont {Shirotori}},
  \bibinfo {author} {\bibfnamefont {P.~D.}\ \bibnamefont {Olmsted}}, \ and\
  \bibinfo {author} {\bibfnamefont {D.}~\bibnamefont {Andelman}},\ }\href
  {http://stacks.iop.org/0295-5075/67/i=2/a=321} {\bibfield  {journal}
  {\bibinfo  {journal} {Europhys. Lett.}\ }\textbf {\bibinfo {volume} {67}},\
  \bibinfo {pages} {321} (\bibinfo {year} {2004})}\BibitemShut {NoStop}%
\bibitem [{\citenamefont {Pantano}\ \emph {et~al.}(2011)\citenamefont
  {Pantano}, \citenamefont {Moore}, \citenamefont {Klein},\ and\ \citenamefont
  {Discher}}]{Pantano2011}%
  \BibitemOpen
  \bibfield  {author} {\bibinfo {author} {\bibfnamefont {D.~A.}\ \bibnamefont
  {Pantano}}, \bibinfo {author} {\bibfnamefont {P.~B.}\ \bibnamefont {Moore}},
  \bibinfo {author} {\bibfnamefont {M.~L.}\ \bibnamefont {Klein}}, \ and\
  \bibinfo {author} {\bibfnamefont {D.~E.}\ \bibnamefont {Discher}},\ }\href
  {\doibase 10.1039/C1SM05490B} {\bibfield  {journal} {\bibinfo  {journal}
  {Soft Matter}\ }\textbf {\bibinfo {volume} {7}},\ \bibinfo {pages} {8182}
  (\bibinfo {year} {2011})}\BibitemShut {NoStop}%
\bibitem [{\citenamefont {Polley}\ \emph {et~al.}(2014)\citenamefont {Polley},
  \citenamefont {Mayor},\ and\ \citenamefont {Rao}}]{Polley2013}%
  \BibitemOpen
  \bibfield  {author} {\bibinfo {author} {\bibfnamefont {A.}~\bibnamefont
  {Polley}}, \bibinfo {author} {\bibfnamefont {S.}~\bibnamefont {Mayor}}, \
  and\ \bibinfo {author} {\bibfnamefont {M.}~\bibnamefont {Rao}},\ }\href
  {\doibase http://dx.doi.org/10.1063/1.4892087} {\bibfield  {journal}
  {\bibinfo  {journal} {J. Chem. Phys.}\ }\textbf {\bibinfo {volume} {141}},\
  \bibinfo {eid} {064903} (\bibinfo {year} {2014})}\BibitemShut {NoStop}%
\bibitem [{\citenamefont {Blosser}\ \emph {et~al.}(2015)\citenamefont
  {Blosser}, \citenamefont {Honerkamp-Smith}, \citenamefont {Han},
  \citenamefont {Haataja},\ and\ \citenamefont {Keller}}]{Blosser2015}%
  \BibitemOpen
  \bibfield  {author} {\bibinfo {author} {\bibfnamefont {M.~C.}\ \bibnamefont
  {Blosser}}, \bibinfo {author} {\bibfnamefont {A.~R.}\ \bibnamefont
  {Honerkamp-Smith}}, \bibinfo {author} {\bibfnamefont {T.}~\bibnamefont
  {Han}}, \bibinfo {author} {\bibfnamefont {M.}~\bibnamefont {Haataja}}, \ and\
  \bibinfo {author} {\bibfnamefont {S.~L.}\ \bibnamefont {Keller}},\ }\href
  {\doibase 10.1016/j.bpj.2014.11.1328} {\bibfield  {journal} {\bibinfo
  {journal} {Biophys. J.}\ }\textbf {\bibinfo {volume} {108}},\ \bibinfo
  {pages} {240a} (\bibinfo {year} {2015})}\BibitemShut {NoStop}%
\bibitem [{\citenamefont {Wallace}(2005)}]{Wallace2005}%
  \BibitemOpen
  \bibfield  {author} {\bibinfo {author} {\bibfnamefont {E.~J.}\ \bibnamefont
  {Wallace}},\ }\emph {\bibinfo {title} {Influence of microstructure on the
  phase behaviour of lipid membranes}},\ \href@noop {} {Ph.D. thesis},\
  \bibinfo  {school} {University of Leeds} (\bibinfo {year} {2005})\BibitemShut
  {NoStop}%
\bibitem [{\citenamefont {Needham}\ and\ \citenamefont
  {Nunn}(1990)}]{Needham1990}%
  \BibitemOpen
  \bibfield  {author} {\bibinfo {author} {\bibfnamefont {D.}~\bibnamefont
  {Needham}}\ and\ \bibinfo {author} {\bibfnamefont {R.}~\bibnamefont {Nunn}},\
  }\href {\doibase http://dx.doi.org/10.1016/S0006-3495(90)82444-9} {\bibfield
  {journal} {\bibinfo  {journal} {Biophys. J.}\ }\textbf {\bibinfo {volume}
  {58}},\ \bibinfo {pages} {997 } (\bibinfo {year} {1990})}\BibitemShut
  {NoStop}%
\bibitem [{\citenamefont {Rawicz}\ \emph {et~al.}(2000)\citenamefont {Rawicz},
  \citenamefont {Olbrich}, \citenamefont {McIntosh}, \citenamefont {Needham},\
  and\ \citenamefont {Evans}}]{Rawicz2000}%
  \BibitemOpen
  \bibfield  {author} {\bibinfo {author} {\bibfnamefont {W.}~\bibnamefont
  {Rawicz}}, \bibinfo {author} {\bibfnamefont {K.}~\bibnamefont {Olbrich}},
  \bibinfo {author} {\bibfnamefont {T.}~\bibnamefont {McIntosh}}, \bibinfo
  {author} {\bibfnamefont {D.}~\bibnamefont {Needham}}, \ and\ \bibinfo
  {author} {\bibfnamefont {E.}~\bibnamefont {Evans}},\ }\href {\doibase
  http://dx.doi.org/10.1016/S0006-3495(00)76295-3} {\bibfield  {journal}
  {\bibinfo  {journal} {Biophys. J.}\ }\textbf {\bibinfo {volume} {79}},\
  \bibinfo {pages} {328 } (\bibinfo {year} {2000})}\BibitemShut {NoStop}%
\bibitem [{\citenamefont {Garc\'{i}a-S\'{a}ez}\ \emph
  {et~al.}(2007)\citenamefont {Garc\'{i}a-S\'{a}ez}, \citenamefont {Chiantia},\
  and\ \citenamefont {Schwille}}]{Garcia2007}%
  \BibitemOpen
  \bibfield  {author} {\bibinfo {author} {\bibfnamefont {A.~J.}\ \bibnamefont
  {Garc\'{i}a-S\'{a}ez}}, \bibinfo {author} {\bibfnamefont {S.}~\bibnamefont
  {Chiantia}}, \ and\ \bibinfo {author} {\bibfnamefont {P.}~\bibnamefont
  {Schwille}},\ }\href {\doibase 10.1074/jbc.M706162200} {\bibfield  {journal}
  {\bibinfo  {journal} {J. Biol. Chem.}\ }\textbf {\bibinfo {volume} {282}},\
  \bibinfo {pages} {33537} (\bibinfo {year} {2007})}\BibitemShut {NoStop}%
\bibitem [{\citenamefont {Lin}\ \emph {et~al.}(2006)\citenamefont {Lin},
  \citenamefont {Blanchette}, \citenamefont {Ratto},\ and\ \citenamefont
  {Longo}}]{Lin2006}%
  \BibitemOpen
  \bibfield  {author} {\bibinfo {author} {\bibfnamefont {W.-C.}\ \bibnamefont
  {Lin}}, \bibinfo {author} {\bibfnamefont {C.~D.}\ \bibnamefont {Blanchette}},
  \bibinfo {author} {\bibfnamefont {T.~V.}\ \bibnamefont {Ratto}}, \ and\
  \bibinfo {author} {\bibfnamefont {M.~L.}\ \bibnamefont {Longo}},\ }\href
  {\doibase http://dx.doi.org/10.1529/biophysj.105.067066} {\bibfield
  {journal} {\bibinfo  {journal} {Biophys. J.}\ }\textbf {\bibinfo {volume}
  {90}},\ \bibinfo {pages} {228 } (\bibinfo {year} {2006})}\BibitemShut
  {NoStop}%
\bibitem [{\citenamefont {Huang}(1987)}]{Huang1987}%
  \BibitemOpen
  \bibfield  {author} {\bibinfo {author} {\bibfnamefont {K.}~\bibnamefont
  {Huang}},\ }\href@noop {} {\emph {\bibinfo {title} {Statistical Mechanics}}}\
  (\bibinfo  {publisher} {Wiley, New York},\ \bibinfo {year}
  {1987})\BibitemShut {NoStop}%
\bibitem [{\citenamefont {Garg}\ \emph {et~al.}(2007)\citenamefont {Garg},
  \citenamefont {R\"{u}he}, \citenamefont {L\"{u}dtke}, \citenamefont
  {Jordan},\ and\ \citenamefont {Naumann}}]{Garg2007}%
  \BibitemOpen
  \bibfield  {author} {\bibinfo {author} {\bibfnamefont {S.}~\bibnamefont
  {Garg}}, \bibinfo {author} {\bibfnamefont {J.}~\bibnamefont {R\"{u}he}},
  \bibinfo {author} {\bibfnamefont {K.}~\bibnamefont {L\"{u}dtke}}, \bibinfo
  {author} {\bibfnamefont {R.}~\bibnamefont {Jordan}}, \ and\ \bibinfo {author}
  {\bibfnamefont {C.~A.}\ \bibnamefont {Naumann}},\ }\href {\doibase
  http://dx.doi.org/10.1529/biophysj.106.091082} {\bibfield  {journal}
  {\bibinfo  {journal} {Biophys. J.}\ }\textbf {\bibinfo {volume} {92}},\
  \bibinfo {pages} {1263 } (\bibinfo {year} {2007})}\BibitemShut {NoStop}%
\bibitem [{\citenamefont {Visco}\ \emph {et~al.}(2014)\citenamefont {Visco},
  \citenamefont {Chiantia},\ and\ \citenamefont {Schwille}}]{Visco2014}%
  \BibitemOpen
  \bibfield  {author} {\bibinfo {author} {\bibfnamefont {I.}~\bibnamefont
  {Visco}}, \bibinfo {author} {\bibfnamefont {S.}~\bibnamefont {Chiantia}}, \
  and\ \bibinfo {author} {\bibfnamefont {P.}~\bibnamefont {Schwille}},\ }\href
  {\doibase 10.1021/la500468r} {\bibfield  {journal} {\bibinfo  {journal}
  {Langmuir}\ }\textbf {\bibinfo {volume} {30}},\ \bibinfo {pages} {7475}
  (\bibinfo {year} {2014})}\BibitemShut {NoStop}%
\bibitem [{\citenamefont {Lin}\ and\ \citenamefont {London}(2015)}]{Lin2015}%
  \BibitemOpen
  \bibfield  {author} {\bibinfo {author} {\bibfnamefont {Q.}~\bibnamefont
  {Lin}}\ and\ \bibinfo {author} {\bibfnamefont {E.}~\bibnamefont {London}},\
  }\href {\doibase 10.1016/j.bpj.2015.03.056} {\bibfield  {journal} {\bibinfo
  {journal} {Biophys. J.}\ }\textbf {\bibinfo {volume} {108}},\ \bibinfo
  {pages} {2212} (\bibinfo {year} {2015})}\BibitemShut {NoStop}%
\bibitem [{\citenamefont {Stukowski}(2010)}]{OVITO}%
  \BibitemOpen
  \bibfield  {author} {\bibinfo {author} {\bibfnamefont {A.}~\bibnamefont
  {Stukowski}},\ }\href {http://stacks.iop.org/0965-0393/18/i=1/a=015012}
  {\bibfield  {journal} {\bibinfo  {journal} {Modelling Simul. Mater. Sci.
  Eng.}\ }\textbf {\bibinfo {volume} {18}},\ \bibinfo {pages} {015012}
  (\bibinfo {year} {2010})}\BibitemShut {NoStop}%
\bibitem [{\citenamefont {Galimzyanov}\ \emph {et~al.}(2015)\citenamefont
  {Galimzyanov}, \citenamefont {Molotkovsky}, \citenamefont {Bozdaganyan},
  \citenamefont {Cohen}, \citenamefont {Pohl},\ and\ \citenamefont
  {Akimov}}]{Galimzyanov2015}%
  \BibitemOpen
  \bibfield  {author} {\bibinfo {author} {\bibfnamefont {T.~R.}\ \bibnamefont
  {Galimzyanov}}, \bibinfo {author} {\bibfnamefont {R.~J.}\ \bibnamefont
  {Molotkovsky}}, \bibinfo {author} {\bibfnamefont {M.~E.}\ \bibnamefont
  {Bozdaganyan}}, \bibinfo {author} {\bibfnamefont {F.~S.}\ \bibnamefont
  {Cohen}}, \bibinfo {author} {\bibfnamefont {P.}~\bibnamefont {Pohl}}, \ and\
  \bibinfo {author} {\bibfnamefont {S.~A.}\ \bibnamefont {Akimov}},\ }\href
  {\doibase 10.1103/PhysRevLett.115.088101} {\bibfield  {journal} {\bibinfo
  {journal} {Phys. Rev. Lett.}\ }\textbf {\bibinfo {volume} {115}},\ \bibinfo
  {pages} {088101} (\bibinfo {year} {2015})}\BibitemShut {NoStop}%
\bibitem [{\citenamefont {Kuzmin}\ \emph {et~al.}(2005)\citenamefont {Kuzmin},
  \citenamefont {Akimov}, \citenamefont {Chizmadzhev}, \citenamefont
  {Zimmerberg},\ and\ \citenamefont {Cohen}}]{Kuzmin2005}%
  \BibitemOpen
  \bibfield  {author} {\bibinfo {author} {\bibfnamefont {P.~I.}\ \bibnamefont
  {Kuzmin}}, \bibinfo {author} {\bibfnamefont {S.~A.}\ \bibnamefont {Akimov}},
  \bibinfo {author} {\bibfnamefont {Y.~A.}\ \bibnamefont {Chizmadzhev}},
  \bibinfo {author} {\bibfnamefont {J.}~\bibnamefont {Zimmerberg}}, \ and\
  \bibinfo {author} {\bibfnamefont {F.~S.}\ \bibnamefont {Cohen}},\ }\href
  {\doibase http://dx.doi.org/10.1529/biophysj.104.048223} {\bibfield
  {journal} {\bibinfo  {journal} {Biophys. J.}\ }\textbf {\bibinfo {volume}
  {88}},\ \bibinfo {pages} {1120 } (\bibinfo {year} {2005})}\BibitemShut
  {NoStop}%
\bibitem [{\citenamefont {Bleecker}\ \emph {et~al.}(2015)\citenamefont
  {Bleecker}, \citenamefont {Cox},\ and\ \citenamefont
  {Keller}}]{Bleecker2015}%
  \BibitemOpen
  \bibfield  {author} {\bibinfo {author} {\bibfnamefont {J.~V.}\ \bibnamefont
  {Bleecker}}, \bibinfo {author} {\bibfnamefont {P.~A.}\ \bibnamefont {Cox}}, \
  and\ \bibinfo {author} {\bibfnamefont {S.~L.}\ \bibnamefont {Keller}},\
  }\href {\doibase 10.1016/j.bpj.2014.11.1333} {\bibfield  {journal} {\bibinfo
  {journal} {Biophys. J.}\ }\textbf {\bibinfo {volume} {108}},\ \bibinfo
  {pages} {241a} (\bibinfo {year} {2015})}\BibitemShut {NoStop}%
\bibitem [{\citenamefont {Jefferys}\ \emph {et~al.}(2014)\citenamefont
  {Jefferys}, \citenamefont {Sansom},\ and\ \citenamefont
  {Fowler}}]{Jefferys2014}%
  \BibitemOpen
  \bibfield  {author} {\bibinfo {author} {\bibfnamefont {E.}~\bibnamefont
  {Jefferys}}, \bibinfo {author} {\bibfnamefont {M.~S.~P.}\ \bibnamefont
  {Sansom}}, \ and\ \bibinfo {author} {\bibfnamefont {P.~W.}\ \bibnamefont
  {Fowler}},\ }\href {\doibase 10.1039/C3FD00131H} {\bibfield  {journal}
  {\bibinfo  {journal} {Faraday Discuss.}\ }\textbf {\bibinfo {volume} {169}},\
  \bibinfo {pages} {209} (\bibinfo {year} {2014})}\BibitemShut {NoStop}%
\bibitem [{Note2()}]{Note2}%
  \BibitemOpen
  \bibinfo {note} {Eq.~\ref {eqn:delEexact} refers to a boundary between R-AR
  domains, whose thickness mismatch is equal to $\Delta _0$, and for large $R$
  the estimated parameters give line tension $\Gamma \approx 2\protect \tmspace
  +\thinmuskip {.1667em}\protect \textrm {pN}$. For an R-R boundary the
  thickness mismatch would double to $2 \Delta _0 \protect \tmspace
  -\thinmuskip {.1667em}\approx \protect \tmspace -\thinmuskip {.1667em}
  2\protect \tmspace +\thinmuskip {.1667em}a \protect \tmspace -\thinmuskip
  {.1667em}\approx \protect \tmspace -\thinmuskip {.1667em} 1.6\protect
  \tmspace +\thinmuskip {.1667em}\protect \textrm {nm}$, yielding line tension
  $\Gamma \approx 8\protect \tmspace +\thinmuskip {.1667em}\protect \textrm
  {pN}$. For this thickness mismatch, the estimate made in \cite {Garcia2007}
  is $\Gamma = 6 \pm 2\protect \tmspace +\thinmuskip {.1667em}\protect \textrm
  {pN}$, using existing theory \cite {Kuzmin2005} which has been shown to yield
  satisfactory agreement with available experiments}\BibitemShut {NoStop}%
\bibitem [{\citenamefont {Ryu}\ and\ \citenamefont {Cai}(2010)}]{Ryu2010}%
  \BibitemOpen
  \bibfield  {author} {\bibinfo {author} {\bibfnamefont {S.}~\bibnamefont
  {Ryu}}\ and\ \bibinfo {author} {\bibfnamefont {W.}~\bibnamefont {Cai}},\
  }\href {\doibase 10.1103/PhysRevE.82.011603} {\bibfield  {journal} {\bibinfo
  {journal} {Phys. Rev. E}\ }\textbf {\bibinfo {volume} {82}},\ \bibinfo
  {pages} {011603} (\bibinfo {year} {2010})}\BibitemShut {NoStop}%
\bibitem [{\citenamefont {Vehkam\"{a}ki}(2006)}]{Vehkamaki2006}%
  \BibitemOpen
  \bibfield  {author} {\bibinfo {author} {\bibfnamefont {H.}~\bibnamefont
  {Vehkam\"{a}ki}},\ }\href@noop {} {\emph {\bibinfo {title} {Classical
  Nucleation Theory in Multicomponent Systems}}}\ (\bibinfo  {publisher}
  {Springer-Verlag},\ \bibinfo {year} {2006})\BibitemShut {NoStop}%
\bibitem [{\citenamefont {Vehkam\"{a}ki}\ \emph {et~al.}(2007)\citenamefont
  {Vehkam\"{a}ki}, \citenamefont {M\"a\"att\"anen}, \citenamefont {Lauri},
  \citenamefont {Napari},\ and\ \citenamefont {Kulmala}}]{Zeldovich2007}%
  \BibitemOpen
  \bibfield  {author} {\bibinfo {author} {\bibfnamefont {H.}~\bibnamefont
  {Vehkam\"{a}ki}}, \bibinfo {author} {\bibfnamefont {A.}~\bibnamefont
  {M\"a\"att\"anen}}, \bibinfo {author} {\bibfnamefont {A.}~\bibnamefont
  {Lauri}}, \bibinfo {author} {\bibfnamefont {I.}~\bibnamefont {Napari}}, \
  and\ \bibinfo {author} {\bibfnamefont {M.}~\bibnamefont {Kulmala}},\ }\href
  {\doibase 10.5194/acp-7-309-2007} {\bibfield  {journal} {\bibinfo  {journal}
  {Atmos. Chem. Phys.}\ }\textbf {\bibinfo {volume} {7}},\ \bibinfo {pages}
  {309} (\bibinfo {year} {2007})}\BibitemShut {NoStop}%
\bibitem [{Note3()}]{Note3}%
  \BibitemOpen
  \bibinfo {note} {Simulations with smaller values of the mismatch parameter
  $\Delta _0$ have been found to yield the same three kinetic classes as
  studied here \cite {Williamson2015}.}\BibitemShut {Stop}%
\bibitem [{\citenamefont {Wallace}\ \emph {et~al.}(2006)\citenamefont
  {Wallace}, \citenamefont {Hooper},\ and\ \citenamefont
  {Olmsted}}]{Wallace2006}%
  \BibitemOpen
  \bibfield  {author} {\bibinfo {author} {\bibfnamefont {E.~J.}\ \bibnamefont
  {Wallace}}, \bibinfo {author} {\bibfnamefont {N.~M.}\ \bibnamefont {Hooper}},
  \ and\ \bibinfo {author} {\bibfnamefont {P.~D.}\ \bibnamefont {Olmsted}},\
  }\href {http://linkinghub.elsevier.com/retrieve/pii/S0006349506725905}
  {\bibfield  {journal} {\bibinfo  {journal} {Biophys. J.}\ }\textbf {\bibinfo
  {volume} {90}},\ \bibinfo {pages} {4104} (\bibinfo {year}
  {2006})}\BibitemShut {NoStop}%
\bibitem [{\citenamefont {Fan}\ \emph {et~al.}(2010)\citenamefont {Fan},
  \citenamefont {Han},\ and\ \citenamefont {Haataja}}]{Fan2010}%
  \BibitemOpen
  \bibfield  {author} {\bibinfo {author} {\bibfnamefont {J.}~\bibnamefont
  {Fan}}, \bibinfo {author} {\bibfnamefont {T.}~\bibnamefont {Han}}, \ and\
  \bibinfo {author} {\bibfnamefont {M.}~\bibnamefont {Haataja}},\ }\href
  {\doibase http://dx.doi.org/10.1063/1.3518458} {\bibfield  {journal}
  {\bibinfo  {journal} {J. Chem. Phys.}\ }\textbf {\bibinfo {volume} {133}},\
  \bibinfo {eid} {235101} (\bibinfo {year} {2010})}\BibitemShut {NoStop}%
\bibitem [{\citenamefont {Korlach}\ \emph {et~al.}(1999)\citenamefont
  {Korlach}, \citenamefont {Schwille}, \citenamefont {Webb},\ and\
  \citenamefont {Feigenson}}]{Korlach1999}%
  \BibitemOpen
  \bibfield  {author} {\bibinfo {author} {\bibfnamefont {J.}~\bibnamefont
  {Korlach}}, \bibinfo {author} {\bibfnamefont {P.}~\bibnamefont {Schwille}},
  \bibinfo {author} {\bibfnamefont {W.~W.}\ \bibnamefont {Webb}}, \ and\
  \bibinfo {author} {\bibfnamefont {G.~W.}\ \bibnamefont {Feigenson}},\ }\href
  {\doibase 10.1073/pnas.96.15.8461} {\bibfield  {journal} {\bibinfo  {journal}
  {Proc. Natl. Acad. Sci.}\ }\textbf {\bibinfo {volume} {96}},\ \bibinfo
  {pages} {8461} (\bibinfo {year} {1999})}\BibitemShut {NoStop}%
\bibitem [{\citenamefont {Dietrich}\ \emph {et~al.}(2001)\citenamefont
  {Dietrich}, \citenamefont {Bagatolli}, \citenamefont {Volovyk}, \citenamefont
  {Thompson}, \citenamefont {Levi}, \citenamefont {Jacobson},\ and\
  \citenamefont {Gratton}}]{Dietrich2001}%
  \BibitemOpen
  \bibfield  {author} {\bibinfo {author} {\bibfnamefont {C.}~\bibnamefont
  {Dietrich}}, \bibinfo {author} {\bibfnamefont {L.}~\bibnamefont {Bagatolli}},
  \bibinfo {author} {\bibfnamefont {Z.}~\bibnamefont {Volovyk}}, \bibinfo
  {author} {\bibfnamefont {N.}~\bibnamefont {Thompson}}, \bibinfo {author}
  {\bibfnamefont {M.}~\bibnamefont {Levi}}, \bibinfo {author} {\bibfnamefont
  {K.}~\bibnamefont {Jacobson}}, \ and\ \bibinfo {author} {\bibfnamefont
  {E.}~\bibnamefont {Gratton}},\ }\href
  {http://linkinghub.elsevier.com/retrieve/pii/S0006349501761140} {\bibfield
  {journal} {\bibinfo  {journal} {Biophys. J.}\ }\textbf {\bibinfo {volume}
  {80}},\ \bibinfo {pages} {1417} (\bibinfo {year} {2001})}\BibitemShut
  {NoStop}%
\bibitem [{\citenamefont {Zhang}\ \emph {et~al.}(2007)\citenamefont {Zhang},
  \citenamefont {Jing}, \citenamefont {Janout},\ and\ \citenamefont
  {Regen}}]{Zhang2007}%
  \BibitemOpen
  \bibfield  {author} {\bibinfo {author} {\bibfnamefont {J.}~\bibnamefont
  {Zhang}}, \bibinfo {author} {\bibfnamefont {B.}~\bibnamefont {Jing}},
  \bibinfo {author} {\bibfnamefont {V.}~\bibnamefont {Janout}}, \ and\ \bibinfo
  {author} {\bibfnamefont {S.~L.}\ \bibnamefont {Regen}},\ }\href {\doibase
  10.1021/la701503v} {\bibfield  {journal} {\bibinfo  {journal} {Langmuir}\
  }\textbf {\bibinfo {volume} {23}},\ \bibinfo {pages} {8709} (\bibinfo {year}
  {2007})}\BibitemShut {NoStop}%
\bibitem [{\citenamefont {Meinhardt}\ \emph {et~al.}(2013)\citenamefont
  {Meinhardt}, \citenamefont {Vink},\ and\ \citenamefont
  {Schmid}}]{Meinhardt2013}%
  \BibitemOpen
  \bibfield  {author} {\bibinfo {author} {\bibfnamefont {S.}~\bibnamefont
  {Meinhardt}}, \bibinfo {author} {\bibfnamefont {R.~L.~C.}\ \bibnamefont
  {Vink}}, \ and\ \bibinfo {author} {\bibfnamefont {F.}~\bibnamefont
  {Schmid}},\ }\href {\doibase 10.1073/pnas.1221075110} {\bibfield  {journal}
  {\bibinfo  {journal} {Proc. Nat. Acad. Sci.}\ }\textbf {\bibinfo {volume}
  {110}},\ \bibinfo {pages} {4476} (\bibinfo {year} {2013})}\BibitemShut
  {NoStop}%
\bibitem [{\citenamefont {Palmieri}\ \emph {et~al.}(2014)\citenamefont
  {Palmieri}, \citenamefont {Yamamoto}, \citenamefont {Brewster},\ and\
  \citenamefont {Safran}}]{Palmieri2014}%
  \BibitemOpen
  \bibfield  {author} {\bibinfo {author} {\bibfnamefont {B.}~\bibnamefont
  {Palmieri}}, \bibinfo {author} {\bibfnamefont {T.}~\bibnamefont {Yamamoto}},
  \bibinfo {author} {\bibfnamefont {R.~C.}\ \bibnamefont {Brewster}}, \ and\
  \bibinfo {author} {\bibfnamefont {S.~A.}\ \bibnamefont {Safran}},\ }\href
  {\doibase http://dx.doi.org/10.1016/j.cis.2014.02.007} {\bibfield  {journal}
  {\bibinfo  {journal} {Adv. Colloid Interface Sci.}\ }\textbf {\bibinfo
  {volume} {208}},\ \bibinfo {pages} {58 } (\bibinfo {year} {2014})},\ \bibinfo
  {note} {special issue in honour of Wolfgang Helfrich}\BibitemShut {NoStop}%
\bibitem [{\citenamefont {Veatch}\ \emph {et~al.}(2007)\citenamefont {Veatch},
  \citenamefont {Soubias}, \citenamefont {Keller},\ and\ \citenamefont
  {Gawrisch}}]{Veatch2007}%
  \BibitemOpen
  \bibfield  {author} {\bibinfo {author} {\bibfnamefont {S.~L.}\ \bibnamefont
  {Veatch}}, \bibinfo {author} {\bibfnamefont {O.}~\bibnamefont {Soubias}},
  \bibinfo {author} {\bibfnamefont {S.~L.}\ \bibnamefont {Keller}}, \ and\
  \bibinfo {author} {\bibfnamefont {K.}~\bibnamefont {Gawrisch}},\ }\href
  {\doibase 10.1073/pnas.0703513104} {\bibfield  {journal} {\bibinfo  {journal}
  {Proc. Natl. Acad. Sci.}\ }\textbf {\bibinfo {volume} {104}},\ \bibinfo
  {pages} {17650} (\bibinfo {year} {2007})}\BibitemShut {NoStop}%
\bibitem [{\citenamefont {Fan}\ \emph {et~al.}(2008)\citenamefont {Fan},
  \citenamefont {Sammalkorpi},\ and\ \citenamefont {Haataja}}]{Fan2008}%
  \BibitemOpen
  \bibfield  {author} {\bibinfo {author} {\bibfnamefont {J.}~\bibnamefont
  {Fan}}, \bibinfo {author} {\bibfnamefont {M.}~\bibnamefont {Sammalkorpi}}, \
  and\ \bibinfo {author} {\bibfnamefont {M.}~\bibnamefont {Haataja}},\ }\href
  {\doibase 10.1103/PhysRevLett.100.178102} {\bibfield  {journal} {\bibinfo
  {journal} {Phys. Rev. Lett.}\ }\textbf {\bibinfo {volume} {100}},\ \bibinfo
  {pages} {178102} (\bibinfo {year} {2008})}\BibitemShut {NoStop}%
\bibitem [{\citenamefont {Turner}\ \emph {et~al.}(2005)\citenamefont {Turner},
  \citenamefont {Sens},\ and\ \citenamefont {Socci}}]{Turner2005}%
  \BibitemOpen
  \bibfield  {author} {\bibinfo {author} {\bibfnamefont {M.~S.}\ \bibnamefont
  {Turner}}, \bibinfo {author} {\bibfnamefont {P.}~\bibnamefont {Sens}}, \ and\
  \bibinfo {author} {\bibfnamefont {N.~D.}\ \bibnamefont {Socci}},\ }\href
  {\doibase 10.1103/PhysRevLett.95.168301} {\bibfield  {journal} {\bibinfo
  {journal} {Phys. Rev. Lett.}\ }\textbf {\bibinfo {volume} {95}},\ \bibinfo
  {pages} {168301} (\bibinfo {year} {2005})}\BibitemShut {NoStop}%
\bibitem [{\citenamefont {Lingwood}\ and\ \citenamefont
  {Simons}(2010)}]{Lingwood2010}%
  \BibitemOpen
  \bibfield  {author} {\bibinfo {author} {\bibfnamefont {D.}~\bibnamefont
  {Lingwood}}\ and\ \bibinfo {author} {\bibfnamefont {K.}~\bibnamefont
  {Simons}},\ }\href {\doibase 10.1126/science.1174621} {\bibfield  {journal}
  {\bibinfo  {journal} {Science}\ }\textbf {\bibinfo {volume} {327}},\ \bibinfo
  {pages} {46} (\bibinfo {year} {2010})}\BibitemShut {NoStop}%
\end{thebibliography}%
\end{document}